\documentclass[12pt]{article}

\usepackage{epsfig}
\usepackage{cite}
\usepackage{amsmath, amssymb, amsfonts}
\usepackage{color}
\usepackage{latexsym}
\usepackage{graphicx}
\usepackage[colorlinks,bookmarks]{hyperref}
\hypersetup{pdfpagemode=UseNone, pdfstartview=FitH, linkcolor=blue,
            citecolor=red, urlcolor=blue}

\def\be{\begin{equation}}
\def\ee{\end{equation}}
\def\ba{\begin{eqnarray}}
\def\ea{\end{eqnarray}}

\def\bdm{\begin{displaymath}}
\def\edm{\end{displaymath}}
\def\la{~\mbox{\raisebox{-.6ex}{$\stackrel{<}{\sim}$}}~}

\def\bq{\begin{quote}}
\def\eq{\end{quote}}

 at 10truept
 at 14truept

\newcommand{\p}{\partial}

\newcommand{\Mpl}{M_{\mathrm{Pl}}}
\newcommand{\mps}{M_{\mathrm{Pl}}^2}

\newcommand{\bea}{\begin{eqnarray}}
\newcommand{\eea}{\end{eqnarray}}

\newcommand{\bi}{\begin{itemize}}
\newcommand{\ei}{\end{itemize}}

\newcommand{\beq}{\begin{equation}}
\newcommand{\eeq}{\end{equation}}
\newcommand{\beqa}{\begin{eqnarray}}
\newcommand{\eeqa}{\end{eqnarray}}
\newcommand{\mpl}{\Mpl}

\def\la{~\mbox{\raisebox{-.6ex}{$\stackrel{<}{\sim}$}}~}

\def\ltap{\ \raise.3ex\hbox{$<$\kern-.75em\lower1ex\hbox{$\sim$}}\ }
\def\gtap{\ \raise.3ex\hbox{$>$\kern-.75em\lower1ex\hbox{$\sim$}}\ }
\def\gl{\ \raise.5ex\hbox{$>$}\kern-.8em\lower.5ex\hbox{$<$}\ }
\def\roughly#1{\raise.3ex\hbox{$#1$\kern-.75em\lower1ex\hbox{$\sim$}}}

\begin{document}

\thispagestyle{empty}
\begin{flushright}
May 2023 \\
\end{flushright}
\vspace*{1.5cm}
\begin{center}

{\Large \bf de Sitter Space Decay and Cosmological Constant}
\vskip.3cm 
{\Large \bf   Relaxation in Unimodular Gravity} 
\vskip.3cm 
{\Large \bf  with Charged Membranes}

\vspace*{1.25cm} {\large 
Nemanja Kaloper$^{a, }$\footnote{\tt
kaloper@physics.ucdavis.edu} 
}\\
\vspace{.2cm} 
{\em $^a$QMAP, Department of Physics and Astronomy, University of
California}\\
\vspace{.05cm}{\em Davis, CA 95616, USA}\\

\vspace{1.7cm} ABSTRACT
\end{center}
General covariant unimodular gravity frameworks, based on the Henneaux-Teitelboim formulation,
are, in disguise, precisely $4$-form field theories corrected with higher dimension operators. In the presence
of charged tensional membranes, any de Sitter space in all such theories is unstable and decays. If the fluxes sourced by membranes are 
mutually incommensurate, de Sitter geometries comprise a very 
refined discretuum of states. Whenever the $4$-form sector is dominated by terms linear in flux the almost-Minkowski space 
is the unique long-time attractor. As a result, a tiny cosmological constant is natural in all such frameworks, without 
appealing to anthropic reasoning. 

\vfill \setcounter{page}{0} \setcounter{footnote}{0}

\vspace{1cm}
\newpage

\vspace{1cm}

\section{Synopsis} 

In this article we demonstrate that general covariant unimodular gravity frameworks, which naturally generalize 
the Henneaux-Teitelboim formulation \cite{Henneaux:1989zc}, 
are nothing other than $4$-form field theories corrected with higher (and lower!) dimension operators, after they are recasted in 
canonically dual variables, analogous to the exchange of coordinates and momenta in classical mechanics. Specifically, the general $4$-form theories 
take exactly the form of generalized unimodular gravity when rewritten in terms of magnetic $4$-form variables, as opposed to the more commonly encountered
electric variables. 

When the theory is completed with the inclusion of
charged tensional membranes, entailing local gauge symmetry, any de Sitter space in all such theories is unstable and decays:  
de Sitter geometries comprise a refined discretuum of unstable states, which decay by nucleation of bubbles bounded 
by the membranes \cite{Brown:1987dd,Brown:1988kg}. 
In the presence of multiple $4$-forms and their associated membrane towers, with at least some of the 
fluxes sourced by charges being incommensurate, the de Sitter discretuum 
is extremely finely grained, with cosmological constant values arbitrarily close to zero. 
Whenever the energy of the $4$-form sector is dominated by leading linear 
terms in $4$-form fluxes, an almost-Minkowski space 
is the unique long-time attractor, in the statistical sense: almost-flat regions greatly outnumber more curved ones \cite{Kaloper:2022oqv,Kaloper:2022utc,Kaloper:2022jpv}.

As a result, a tiny cosmological constant 
becomes natural in all such frameworks, without appealing to anthropic reasoning. We discuss how this can be employed to address
the cosmological constant problem in our universe, and various aspects of cosmology.

{%\bf 
In the interest of clarity, let us stress again that our main new results here are 
\begin{itemize}
\item establishing the 1-to-1 correspondence between generic $4$-form theories
and a dynamical generalization of unimodular gravity, which includes charged membranes which source the fluxes that screen the 
cosmological constant; in this vein, we find that the standard covariantly formulated unimodular gravity \cite{Henneaux:1989zc} is
interpreted as a Routhian transform of the $4$-form theories; this changes the perspective on the theory and should
affect how the theory is UV-completed and quantized;
\item demonstrating that for generic $4$-form fluxes, the presence of at least two systems of membranes which source incommensurate 
contributions to the cosmological constant such that the terms linear in fluxes \underline{dominate} over higher powers,
the attractor mechanism discovered and elaborated in \cite{Kaloper:2022oqv,Kaloper:2022utc,Kaloper:2022jpv} remains
fully operational;
\item confirming that quantum dynamics statistically biases the distribution of vacua, dynamically exponentially favoring $\Lambda \rightarrow 0^+$;
other values of $\Lambda$ are possible, but dynamically suppressed\footnote{Although \cite{Kaloper:2022oqv,Kaloper:2022utc,Kaloper:2022jpv} focus
only on linear flux terms, it was noted that more general examples will behave in the same manner, which we show in detail here.}, since 
the linear flux terms, when dominant, allow only processes
which have rates $\approx \exp(-24\pi^2 \mpl^4/\Lambda)$; this is a loophole around 
the more traditional approaches to flux generated landscapes which have nearly-uniform distribution of vacua. 
\end{itemize}
}

\section{What is General Relativity?}

In 1915, Einstein \cite{Einstein:1915ca}  and Hilbert \cite{Hilbert:1915tx} laid foundations of General Relativity (GR). The former wrote
the field equations governing geometry sourced by stress energy of dynamical matter. The latter formulated the simplest action
principle which the field equations can be derived from. Enforcing locality and causality by imposing on the theory to involve no more
than two derivatives, and encoding local gauge invariance via Bianchi's identities, dimensional analysis requires that gravity's coupling to matter 
is a dimensional constant, identified with Newton's constant $G_N = 1/8\pi \mpl^2$, where $\mpl$ is Planck mass. 
Another dimensional constant $\Lambda$ arises representing the energy of the vacuum 
of the theory, which not only doesn't need to be zero, but in quantum theory most likely can't be, thanks to Equivalence Principle and quantum 
uncertainty, as originally noted by Bronstein \cite{Volovik:2007fi} and Pauli \cite{ruzi}, and formulated in more modern terms by 
Zeldovich \cite{Zeldovich:1967gd}. Since the engineering dimension of $\Lambda$ 
is four, it is often viewed as a quartic power of a UV cutoff scale (or some symmetry breaking
scale, such as SUSY and/or conformal symmetry). Yet, since it sources gravity via Einstein's equations, it's effect on the geometry
is via inducing a vacuum curvature ${\cal H}^2 = \Lambda/3 \mpl^2 $. Thus inverting this, $\Lambda$ can be viewed as a
square of an IR cutoff set by curvature, since at longer distances bending of space cannot be ignored. 

Since both of these terms are constants it is tempting -- and common -- to set them to their observed values from the get-go, and treat them as given dimensional
parameters of the theory, in both the field equations and the action. Yet there is absolutely no {\it \`a priori} argument to select their numerical values from first principles. 
On the contrary, the hugely discrepant scales controlling these numbers, and additionally their disparity with any of the known scales in the matter sector,
are commonly recognized as the gauge hierarchy and cosmological constant problems, respectively (see, e.g., \cite{Wilczek:1983as,Weinberg:1987dv}). 
However, a careful examination of quantum field theory (QFT) coupled to semiclassical gravity, specifically renormalization and UV sensitivity of theory's
operator expansion, clearly shows that in this limit there is no reason to expect these numbers to be correlated \cite{Kaloper:2014dqa}. Both
are UV sensitive, and after renormalization, they both depend on independent finite parts of the bare counterterms added to cancel divergences. 
This, along with the na\"ive idea of naturalness in effective field theory (EFT), suggests that the cosmological constant problem, which can be loosely phrased as
$$\Bigl(\frac{\Lambda}{\mpl^4}\Bigr) \Bigl|_{\tt observed} \ll 1 ~~~~~ {\rm versus}  ~~~~~ \Bigl(\frac{\Lambda}{\mpl^4}\Bigr) \Bigl|_{\tt calculated} \gg \Bigl(\frac{\Lambda}{\mpl^4}\Bigr) \Bigl|_{\tt observed} \, ,$$
is a ``radiative instability": quantum corrections want to drive $\Lambda$ up. Yet, without a
UV completion including gravity, this issue is moot, since in generic cases renormalization introduces a separate 
counterterm for each UV sensitive quantity with totally arbitrary finite parts\footnote{Symmetries can correlate them, but an example of such symmerties 
which fits our Universe has not been found to date.}. 
Thus in a QFT, $\Bigl(\frac{\Lambda}{\mpl^4}\Bigr) \Bigl|_{\tt calculated}$ can be {\it anything}. A generic theory doesn't fail to fix the counterterms, 
it simply isn't supposed to. 
That doesn't mean that there's no cosmological constant problem. We will discuss this subtle issue in more detail in what follows. 

From the point of view of `canonical' GR based on Einstein-Hilbert theory \cite{Einstein:1915ca,Hilbert:1915tx}, this is particularly jarring. 
After all, the canonical action does not even seem to have the degrees of freedom that could accommodate variation of $\Lambda$, and the many attempts 
to append them to the theory in the EFT framework have run into the wall of venerable Weinberg's no-go theorem \cite{Weinberg:1987dv} (see also \cite{Kaloper:2014dqa}).
Yet, only a few years after the paper on formulating GR, Einstein proposed its, perhaps most minimal, generalization  \cite{Einstein:1919gv}, which today 
is commonly dubbed {\it unimodular} gravity (see, e.g. 
\cite{Anderson:1971pn,Unruh:1988in,Buchmuller:1988wx,Buchmuller:1988yn,Henneaux:1989zc,Ng:1990xz,Fiol:2008vk,deBrito:2021pmw,Kugo:2021bej}). 
In a nutshell, in this approach the cosmological parameter $\Lambda$ is viewed as yet another free parameter of the theory, which solves a field equation, albeit trivially:
$\partial_\mu \Lambda = 0$, and hence it is an arbitrary constant. This follows from gauge invariance of the theory: the local contributions to the metric determinant $\det(g_{\mu\nu})$ are pure gauge, since all
local propagating metric modes cancel out of it. It can be set to any value by a diffeomorphism, under which 
$\det(g_{\mu\nu}) \rightarrow \det(g'_{\mu\nu}) = \det(g_{\mu\nu}) [\det(\partial x/\partial x')]^2$.
In particular one could choose the gauge $\det(g_{\mu\nu})=1$. This gauge-fixes the diffeomorphism transformations 
to only those with $|\det(\partial x/\partial x')| = 1$ (hence the moniker ``unimodular"\footnote{This notion of unimodularity is somewhat of a distraction, since gauge fixing 
can be altered at will; the one global gauge invariant degree of freedom in $\int d^4x \sqrt{g}$ is completely unaffected by it, as we discuss in what follows.}). Since
it is now fixed, one drops the field        
equation obtained by $\det(g_{\mu\nu})$ variation. This amounts to retaining only the traceless subset
of Einstein's equations. However, Bianchi identities, which follow from local gauge invariance, recover the trace, albeit under a derivative \cite{Weinberg:1987dv}.
The effective theory is split into $9+1$ field equations,
\be
\mpl^2 \Bigl(R^\mu{}_\nu - \frac14 R \delta^\mu{}_\nu \Bigr) = - \Bigl(T^\mu{}_\nu - \frac14 T \delta^\mu{}_\nu\Bigr) \, , ~~~~~~~~ \partial_\mu \Bigl(\mpl^2 R - T \Bigr) = 0 \, .
\label{unimo}
\ee
Integrating the last equation, and plugging the result back into the 
traceless field equations yields the ``rearranged" full set of $10$ covariant equations
valid in any gauge
\be
\mpl^2 \Bigl(R^\mu{}_\nu - \frac12 R \delta^\mu{}_\nu  \Bigr) = - T^\mu{}_\nu -  \Lambda \delta^\mu{}_\nu\, ,
\label{grcc}
\ee
where $\Lambda$ is the integration constant arising from $\partial_\mu \Bigl(\mpl^2 R -T \Bigr) = 0$ (normalized to dim. 4). Thus the only distinction between 
canonical GR with fixed $G_N$ and $\Lambda$, and unimodular GR is the complete arbitrariness of $\Lambda$ in the latter interpretation,
which is actually more aligned with the description of renormalized QFT dwelling on the background spacetime. The integration constant $\Lambda$ 
is precisely the required QFT counterterm which subtracts the UV sensitive contribution from QFT \cite{Kaloper:2014dqa}. While this feature motivated some claims that 
as a result cosmological constant problem is solved in unimodular GR, that is manifestly {\it not true}  \cite{Weinberg:1987dv}. The physical cosmological constant which 
bends the spacetime vacuum $| 0 \rangle$ is\footnote{The vacuum expectation $\langle 0 | T^\mu{}_\mu | 0 \rangle$  projects $T^\mu{}_\mu$ to only its constant part.}
\be
\Lambda_{\tt phys} = \Lambda + \frac14 \langle 0 | T^\mu{}_\mu | 0 \rangle \, ,
\label{lphys}
\ee
and so $\Lambda$ can indeed be chosen such that $\Lambda_{\tt phys}$ is finite, but its value, without additional 
ingredients (see, e.g. \cite{Kaloper:2014dqa,Kaloper:2013zca,Kaloper:2015jra}), remains indeterminate (see also Sec. VII of \cite{Weinberg:1987dv}). 

This one parameter freedom arises because $\det(g_{\mu\nu})$ does depend on a {\it single} global gauge invariant degree of freedom: the spacetime volume
$\Omega_4 = \int d^4x \sqrt{\det(g_{\mu\nu})|}$ is gauge invariant (although often ill defined due to infinities) 
\cite{Kaloper:2014dqa,Kaloper:2013zca,Kaloper:2015jra,deBrito:2021pmw}.
Nevertheless, its Legendre dual is precisely $\Lambda$, which can be treated as a regulated, gauge invariant, global degree of freedom, of so-generalized GR. 
 
Formulating the correct gauge invariant action principle for unimodular gravity however took a while. 
Gauge-fixing $g = \det(g_{\mu\nu})=1$ via a Lagrange multiplier in Einstein-Hilbert action,
\be
S = \int d^4x \Bigl\{\sqrt{g} \Bigl( \frac{\mpl^2}{2} R  - {\cal L}_{\tt QFT} \Bigr) - \Lambda \Bigl(\sqrt{g} -1\Bigr)\Bigr\} \, ,
\label{badaction}
\ee
and treating $\Lambda$ as a global variable (set to a constant to satisfy Bianchi identities) reproduces (\ref{unimo}) at the classical level. However, the very last term violates
gauge symmetry (i.e. diffeomorphism invariance) and so, by itself, it could compromise the gauge symmetry which enforces $\Lambda = {\rm const}.$ in full quantum theory. 
However the manifestly covariant formulation given by Henneaux and Teitelboim in \cite{Henneaux:1989zc} precludes any such concerns. With our normalizations,
the covariant unimodular action of Henneaux and Teitelboim 
is\footnote{In \cite{Henneaux:1989zc}, the authors mostly use the vector Hodge dual to the $3$-form potential, 
${\cal A}_{\mu\nu\lambda} = \sqrt{g} \epsilon_{\mu\nu\lambda\sigma} {\cal T}^\sigma$,
but do identify the dual relationship explicitly in their Eqs. (23)-(26).}
\be
S = \int d^4x \Bigl\{\sqrt{g} \Bigl(\frac{\mpl^2}{2} R  - {\cal L}_{\tt QFT} - \Lambda   \Bigr) +
\frac{1}{3 \mpl^2} {\epsilon^{\mu\nu\lambda\sigma}} \Bigl(\partial_\mu \Lambda \Bigr){\cal A}_{\nu\lambda\sigma}
 \Bigr\} \, .
 \label{actionht} 
\ee
Here ${\cal A}_{\nu\lambda\sigma}$ is an auxiliary $3$-form gauge potential which serves as a Lagrange multiplier. 
Variation of this action with respect to the metric yields the standard GR equations with a ``free function" $\Lambda$, which is constrained to
a constant by the field equation $\partial_\mu \Lambda = 0$ arising from ${\cal A}_{\nu\lambda\sigma}$ variation. Finally, variation with respect to
$\Lambda$ itself yields a ``spectator"  equation ${\epsilon^{\mu\nu\lambda\sigma}} \partial_\mu {\cal A}_{\nu\lambda\sigma} = -{3 \mpl^2} \sqrt{g}$.
If we introduce a ``spectator" $4$-form ${\cal F}_{\mu\nu\lambda\sigma} = 4 \partial_{[\mu} {\cal A}_{\nu\lambda\sigma]}$ (where $[\ldots]$ denotes 
antisymmetrization) we can invert this equation to ${\cal F}_{\mu\nu\lambda\sigma} = \frac{\mpl^2}{2} \sqrt{g}\epsilon_{\mu\nu\lambda\sigma}$.

Note, that this $4$-form flux does not gravitate directly since it is completely decoupled from the metric. Yet, it points to an interesting link between
unimodular gravity and theories with $4$-forms, already noted in Sec. VII of \cite{Weinberg:1987dv}, but otherwise being an apparently as yet untold story.
We will unveil this link in detail in the next section, after we give a straightforward generalization of the Henneaux-Teitelboim action (\ref{actionht}). But
first, let us remark that it has been proven that the action (\ref{actionht}), which is clearly classically identical to conventional GR in every respect except that
$\Lambda$ is an arbitrary integration constant, preserves this identity in quantum theory, both in the semiclassical limit \cite{Fiol:2008vk} 
and in the loop expansion as a QFT \cite{deBrito:2021pmw,Kugo:2021bej}. 

To conclude this section, we first slightly rewrite (\ref{actionht}), by redefining $\Lambda = \mpl^2 \lambda$; the action becomes
\be
S = \int d^4x \Bigl\{\sqrt{g} \Bigl(\frac{\mpl^2}{2} R  - \mpl^2 \lambda - {\cal L}_{\tt QFT} \Bigr) +
\frac{1}{3} {\epsilon^{\mu\nu\lambda\sigma}} \Bigl(\partial_\mu \lambda \Bigr){\cal A}_{\nu\lambda\sigma}
 \Bigr\} \, .
 \label{actionhtr} 
\ee
One can readily verify that this action is identical to a subset of actions recently discussed in \cite{Kaloper:2022oqv,Kaloper:2022utc,Kaloper:2022jpv}, in connection
to the cosmological constant relaxation toward zero (neglecting for the moment the boundary terms in the
actions in \cite{Kaloper:2022oqv,Kaloper:2022utc,Kaloper:2022jpv}). 
Now, viewing $\lambda$ as the independent variable, it immediately leaps to the eye that the `bulk' term 
$\mpl^2 \lambda$ could be viewed as a truncation of some more general `Lagrangian' ${\cal L}(\lambda)$ to the leading term in Taylor expansion. 
Indeed, if one is to compute the cosmological constant contributions from ${\cal L}_{\tt QFT}$ using some regulator $\mu$, one will find
that the loop expansion yields $\Lambda_{\tt QFT} = c_0 \mu^4 + c_1 \mu^2 m^2 + c_2 m^4 + c_3 m^6/\mu^2 + \ldots$, up to logarithms of the
regulator $\mu$, where $m$ is the mass describing local QFT degrees of freedom. In this instance, one can take $\mu = \mpl$ in order to
include graviton loops as well as those from the QFT. 

It is thus only natural to generalize (\ref{actionht}) to 
\be
S = \int d^4x \Bigl\{\sqrt{g} \Bigl(\frac{\mpl^2}{2} R  - {\cal L}(\lambda) - {\cal L}_{\tt QFT} \Bigr) +
\frac{1}{3} {\epsilon^{\mu\nu\lambda\sigma}} \Bigl(\partial_\mu \lambda \Bigr){\cal A}_{\nu\lambda\sigma}
 \Bigr\} \, ,
 \label{actionhtgen} 
\ee
where\footnote{We ignore $c_0 \mpl^4$ in the expansion since that term can be absorbed away by a finite renormalization of $\Lambda_{\tt QFT}$. Normalizing powers of $2$ are for latter
convenience.}    
\be
{\cal L}(\lambda) = \mpl^4 \Bigl(c_1 \frac{\lambda}{ \mpl^2 } + 2 c_2 \frac{\lambda^2}{\mpl^4} + 4 c_3 \frac{\lambda^3}{\mpl^6} + \ldots \Bigr) \, .
\label{lambdalag}
\ee
At the classical level and without charges, however, this generalization is trivial, since it is merely replacing the integration constant $\mpl^2 \lambda$ by a more
general function ${\cal L}(\lambda)$ in the gravitational field equations, which can be undone with a field redefinition\footnote{To be discussed in more detail shortly.}. 
However, when we transition to the dual variables, and add charged discrete degrees of freedom
that can change $\lambda$, the real purpose of this generalization will become manifest. We now turn to these issues.

\section{Unimodular Gravity in Dual Variables}

To illustrate the dualization procedure, let us consider a simple harmonic oscillator, with the Hamiltonian 
$H = p^2/2 + q^2/2$. Clearly, $p$ and $q$ are canonical variables, whose Hamilton's equations are
$\dot q = \partial_p H = p$ and $\dot p = - \partial_q H = - q$. The equivalent $2^{nd}$ order equation is
$\ddot q + q = 0$. However, consider now the change of variables $P = q, Q = -p$. In terms of those
variables, the Hamiltonian is $H = P^2/2 + Q^2/2$, and the equations of motion retain the same structure as
before: $\dot Q = \partial_P H$, $\dot P = - \partial_Q H$. Indeed, as is well known, the symplectic transformations of
generalized coordinates are dynamical symmetries, and in our example, comprise a canonical transformation
of the theory. Either set of variables is as good a description, since no information about the dynamics is neither
``lost" nor ``found". The only question is, which of the variables is more convenient to address a specific application
of the theory. 

The same is true for other dynamical structures. Given a dynamical system with its set of
generalized coordinates and momenta, we are free to rename and reshuffle variables at will provided that
we preserve the canonical structure of the theory -- i.e. that the transformations are canonical in the sense
of Hamiltonian mechanics. We are particularly interested in applying such transformations to the theories 
involving $4$-forms, which have a long history in the pursuit of mechanisms to address the cosmological constant problem \cite{Aurilia:1980xj,Duff:1980qv,Hawking:1981gd,Baum:1983iwr,Hawking:1984hk,Brown:1987dd,Brown:1988kg,Duff:1989ah,Duncan:1989ug,Duncan:1990fr,Aurilia:1993xi,Klinkhamer:2007pe,Bousso:2000xa,Feng:2000if,Dvali:2001sm,Dvali:2003br,Dvali:2005an}. We look below at a few examples and 
eventually at the general case.

\subsection{\it Quadratic 4-form}

A most common example encountered in the past works is a theory given by the bulk action 
\be
S = \int d^4x \sqrt{g} \Bigl(\frac{\mpl^2}{2} R  - {\cal L}_{\tt QFT} - \frac{1}{48} {\cal F}_{\mu\nu\lambda\sigma}^2 \Bigr) \, ,
 \label{actionforms} 
\ee
where ${\cal F}_{\mu\nu\lambda\sigma} = 4 \partial_{[\mu} {\cal A}_{\nu\lambda\sigma]}$ for some $3$-form ${\cal A}$, ignoring for the moment 
the membranes charged under ${\cal A}$. Rather than discussing the field equations, we cut to the chase and dualize this
theory using the standard techniques explained in, e.g. \cite{Dvali:2005an,ks1,ks2,ks3}. That trick will make the connection of
(\ref{actionforms}) to unimodular gravity quickly and clearly. The idea is to reformulate  (\ref{actionforms}) as a first-order theory,
by adding the Lagrangian constraint $\frac{1}{12}\lambda \epsilon^{\mu\nu\lambda\sigma}({\cal F}_{\mu\nu\lambda\sigma} - 4 \partial_{\mu} {\cal A}_{\nu\lambda\sigma})$,
and then integrating out ${\cal F}$. This is most simply illustrated with the path integral
\be
Z = \int \ldots [{\cal D} {\cal A}] [{\cal D}{\cal F}]   [{\cal D}\lambda] \, e^{i S +  2 i \int  \lambda ( {\cal F} -d {\cal A})} \ldots \, ,
\label{partf} 
\ee
where the total action is (with a boundary term to be added below)
\ba
S_{\rm total} &=& S + 2 \int  \lambda ( {\cal F} -d {\cal A})  \nonumber \\ 
&=& \int d^4x \Bigl\{\sqrt{g} \Bigl(\frac{\mpl^2}{2} R  - {\cal L}_{\tt QFT} - \frac{1}{48} {\cal F}_{\mu\nu\lambda\sigma}^2 \Bigr) 
+ \frac{1}{12}\lambda \epsilon^{\mu\nu\lambda\sigma}
\bigl({\cal F}_{\mu\nu\lambda\sigma} - 4 \partial_{\mu} {\cal A}_{\nu\lambda\sigma} \bigr)\Bigr\} \, .
\label{firstorder}
\ea
Defining a new independent degree of freedom
\be
\tilde {\cal F}_{\mu\nu\lambda\sigma} = {\cal F}_{\mu\nu\lambda\sigma} - 2 \lambda {\sqrt{g}} {\epsilon_{\mu\nu\lambda\sigma}} \, ,
\label{trans}
\ee 
and recalling that the translation of variables as in (\ref{trans}) do not change 
the path integral measure since the functional
Jacobian is unity, we rewrite the action ({\it sans} the subscript),   
\be
S = \int d^4x \Bigl\{\sqrt{g} \Bigl(\frac{\mpl^2}{2} R  - {\cal L}_{\tt QFT} - {2} \lambda^2 - \frac14 \tilde {\cal F}_{\mu\nu\lambda\sigma}^2 \Bigr) 
- \frac{1}{3}\lambda \epsilon^{\mu\nu\lambda\sigma} \partial_{\mu} {\cal A}_{\nu\lambda\sigma} \Bigr\}  \, . 
\label{firstorder2}
\ee
Since $\tilde {\cal F}$ does not appear anywhere else, the integration over it yields a factorizable Gaussian normalization factor,
\be
Z = \int \ldots [{\cal D}\tilde {\cal F}]  \dots e^{\ldots + i \int d^4x\sqrt{g} \Bigl( - \frac{1}{48} \tilde {\cal F}_{\mu\nu\lambda\sigma}^2
 \Bigr) } \ldots \, ,
\ee
which can be dropped. Then adding the boundary term for the $4$-form sector, required by 
intrinsic consistency of the variational principle \cite{Duncan:1989ug,ks1},
which is just $\int \frac{1}{3} \epsilon^{\mu\nu\lambda\sigma} \partial_{\mu} (\lambda \, {\cal A}_{\nu\lambda\sigma})$, our action, 
in terms of the new dual variable $\lambda$, becomes 
\be
S = \int d^4x \Bigl\{\sqrt{g} \Bigl(\frac{\mpl^2}{2} R  - {\cal L}_{\tt QFT} - {2} \lambda^2 \Bigr) 
+ \frac{1}{3} \epsilon^{\mu\nu\lambda\sigma} \partial_\mu \Bigl(\lambda\Bigr)  {\cal A}_{\nu\lambda\sigma} \Bigr\}  \, .
\label{actionnewmemd} 
\ee
Note that the signs came up as they do due to $\epsilon_{\mu\nu\lambda\sigma}\epsilon^{\mu\nu\lambda\sigma} = - 4!$ in 
Lorentzian signature. 

\underline{This is precisely the action (\ref{actionhtgen}) with ${\cal L}(\lambda) = 2 \lambda^2$}. The actual normalizing 
factor of $\lambda^2$ does not matter when we ignore the membranes charged under ${\cal A}$, since it can be set to an arbitrary value
by rescaling $\lambda$ and ${\cal A}$ (with charges present, this involves a finite rescaling of charges as well). 
However, the important point is that the bulk 4-form theory is nothing other than a variant of 
unimodular gravity without charges. It has been noticed by Weinberg that 4-form flux energy 
behaves like a unimodular gravity cosmological counterterm (see Sec VII of [10]), but in 
fact we will argue here that this connection is more than just an analogy. The precise map linking these pictures  
is that a theory with $4$-forms and tensional membranes becomes a unimodular gravity in the decoupling limit, when the membrane tension
is taken to infinity. 

\subsection{\it Linear + Quadratic}

It is obvious that these statements are true for more general examples of ${\cal L}(\lambda)$, or in other words for actions which include a variety
of powers of $F_{\mu\nu\lambda\sigma}$. The dualization procedure is more complicated, but it amounts to performing a dual transormation
${\cal F} \leftrightarrow  {{^\ast}\lambda}$, and replacing $\hat {\cal L}({\cal F})$ by its Legendre transform ${\cal L}(\lambda)$  \cite{Dvali:2005an,Klinkhamer:2007pe,Kaloper:2020jso}. 
As an illustration, let's first dualize 
\be
S = \int d^4x \Bigl\{ \sqrt{g} \Bigl(\frac{\mpl^2}{2} R  - {\cal L}_{\tt QFT} - \frac{1}{48} {\cal F}_{\mu\nu\lambda\sigma}^2 \Bigr)
- \frac{\alpha}{24} \epsilon^{\mu\nu\lambda\sigma} {\cal F}_{\mu\nu\lambda\sigma} \Bigr\} \, .
 \label{linsquare} 
\ee
Such a combination of $4$-form terms can be found in, e.g. \cite{Aurilia:1980xj}. 
Here, $\alpha$ is a fixed $4$-form theory coupling parameter which can be induced by nontrivial CP-breaking effects
\cite{Aurilia:1980xj}. Following the same steps as outlined above (and adding 
the boundary term $\int \frac{1}{3} \epsilon^{\mu\nu\lambda\sigma} \partial_{\mu} (\lambda \, {\cal A}_{\nu\lambda\sigma})$ 
to (\ref{linsquare}))
\be
S = \int \Bigl\{ \sqrt{g} \Bigl(\frac{\mpl^2}{2} R  - {\cal L}_{\tt QFT} - \frac{1}{48} {\cal F}_{\mu\nu\lambda\sigma}^2 \Bigr) - \frac{\alpha}{24} \epsilon^{\mu\nu\lambda\sigma} {\cal F}_{\mu\nu\lambda\sigma}
+ \frac{1}{12}\lambda \epsilon^{\mu\nu\lambda\sigma} {\cal F}_{\mu\nu\lambda\sigma} + \frac{1}{3} \epsilon^{\mu\nu\lambda\sigma} 
\partial_{\mu} \Bigl( \lambda \Bigr) {\cal A}_{\nu\lambda\sigma} \bigr)
 \Bigr\} \, . 
 \label{linsquare2} 
\ee
Defining a new variable $\tilde {\cal F}_{\mu\nu\lambda\sigma} = {\cal F}_{\mu\nu\lambda\sigma} - (2 \lambda - \alpha) {\sqrt{g}} {\epsilon_{\mu\nu\lambda\sigma}} $ 
and integrating out $\tilde{\cal F}$, after a straightforward algebra we find
\be
S = \int d^4x \Bigl\{\sqrt{g} \Bigl(\frac{\mpl^2}{2} R  - {\cal L}_{\tt QFT} - {2} \bigl(\lambda-\frac{\alpha}{2})^2 \Bigr) 
+ \frac{1}{3} \epsilon^{\mu\nu\lambda\sigma} \partial_\mu \Bigl(\lambda\Bigr)  {\cal A}_{\nu\lambda\sigma} \Bigr\}  \, .
\label{linsquare3}  
\ee
Note that ${\cal L}(\lambda) = {2} \bigl(\lambda-\frac{\alpha}{2})^2$ is precisely the Legendre transform of 
$\hat {\cal L}({\cal F})= \frac{1}{48} {\cal F}_{\mu\nu\lambda\sigma}^2 + \frac{\alpha}{24 \sqrt{g}} \epsilon^{\mu\nu\lambda\sigma} {\cal F}_{\mu\nu\lambda\sigma}$. 
Indeed, defining the variable\footnote{This shows that the odd powers of ${\cal F}$ in the action may be viewed as integrals over an alternative measure
${\cal F } = \frac{1}{4!}{\cal F}_{\mu\nu\lambda\sigma} dx^\mu dx^\nu dx^\lambda dx^\sigma$, as opposed to $\sqrt{g} d^4x$, as noted in  
\cite{Kaloper:2022oqv,Kaloper:2022utc} and also in e.g.\cite{Unruh:1988in,Guendelman:1996qy,Gronwald:1997ei,Wilczek:1998ea}.} 
${\cal F} =  2 {\bigl(\frac{1}{4! \sqrt{g}} \epsilon^{\mu\nu\lambda\sigma} {\cal F}_{\mu\nu\lambda\sigma}\bigr)}$ 
and using the identity ${\cal F}_{\mu\nu\lambda\sigma}^2 = - 3! {\cal F}^2$, 
it follows that $\lambda$ and ${\cal F}$ are related by 
\be
\lambda \equiv   \frac{ \partial \hat {\cal L}({\cal F})}{\partial {\cal F}} 
= - \frac{\cal F}{4} + \frac12 \alpha \, ,
\label{leg1}
\ee
and solving it to eliminate ${\cal F} = 2(\alpha - 2\lambda)$, we find 
the Legendre transform of ${\cal L}({\cal F})$ to be
\be
{\cal L}({\lambda}) = \hat {\cal L}({\cal F})  - \lambda {\cal F} \equiv  {2} \bigl(\lambda-\frac{\alpha}{2})^2 \, .
\label{legtransf}
\ee

\subsection{\it General Unimodular Case: Multiple Powers of the 4-form}

The same procedure works for a general $\hat {\cal L}({\cal F})$. In Eq. (\ref{linsquare2}), all we do is replace 
$\frac{1}{48} {\cal F}_{\mu\nu\lambda\sigma}^2 + \frac{\alpha}{24 \sqrt{g}} \epsilon^{\mu\nu\lambda\sigma} {\cal F}_{\mu\nu\lambda\sigma}$ 
with $\hat {\cal L}({\cal F})$, and define the
variable $\tilde {\cal F}_{\mu\nu\lambda\sigma} = {\cal F}_{\mu\nu\lambda\sigma} - 2 \lambda {\sqrt{g}} {\epsilon_{\mu\nu\lambda\sigma}} $ 
which is guaranteed to gather all the
${\cal F}$-dependent pieces (and produce the leftover $\propto \lambda$ terms), which can then be integrated out in the path integral. 
What remains is the dual action for the new variable $\lambda$, which will take the form of
(\ref{linsquare3}), with ${2} \bigl(\lambda-\frac{\alpha}{2})^2$ replaced by the Legendre transform of $\hat {\cal L}({\cal F})$: defining the variable 
${\cal F} =  2 {\bigl(\frac{1}{4! \sqrt{g}} \epsilon^{\mu\nu\lambda\sigma} {\cal F}_{\mu\nu\lambda\sigma}\bigr)}$, one can show that 
\be
\lambda \equiv  \frac{ \partial \hat {\cal L}({\cal F})}{\partial {\cal F}} \, , ~~~~~~~~~~ {\cal L}({\lambda}) = \hat {\cal L}({\cal F})  - \lambda {\cal F} \, .
\label{legtransfgen}
\ee
Thus the magnetic dual representation of {\it any} $4$-form theory is precisely 
the general form of unimodular gravity action introduced in the previous section. We repeat it here for clarity: 
\be
S = \int d^4x \Bigl\{\sqrt{g} \Bigl(\frac{\mpl^2}{2} R  - {\cal L}_{\tt QFT} - {\cal L}(\lambda) \Bigr) 
+ \frac{1}{3} \epsilon^{\mu\nu\lambda\sigma} \partial_\mu \Bigl(\lambda\Bigr)  {\cal A}_{\nu\lambda\sigma} \Bigr\}  \, .
\label{genuni} 
\ee
Obviously, this procedure is in general invertible, and so starting with the theory (\ref{genuni}), we can rewrite it as a theory of a $4$-form
\be
S = \int \Bigl\{ \sqrt{g} \Bigl(\frac{\mpl^2}{2} R  - {\cal L}_{\tt QFT} - \hat {\cal L}({\cal F}) \Bigr) + \frac{1}{12}\lambda \epsilon^{\mu\nu\lambda\sigma}
\bigl({\cal F}_{\mu\nu\lambda\sigma} - 4 \partial_{\mu} {\cal A}_{\nu\lambda\sigma} \bigr)
 + \frac{1}{3} \epsilon^{\mu\nu\lambda\sigma} \partial_{\mu} (\lambda \, {\cal A}_{\nu\lambda\sigma})  \Bigr\} \, . 
 \label{gen4form} 
\ee 
We stress again, that despite appearances (\ref{genuni}) and (\ref{gen4form}) \underline{represent the same theory}. 
Note also that in all cases of interest to us we have assumed that the bulk action only depended on ${\cal F} = d{\cal A}$, which means
that ${\cal A}$ is the field theory analogue of cyclic coordinates in mechanics. This means that our procedure of trading the
`conserved' quantity ${\cal F}$ for its `integral of motion' $\lambda$ is really the procedure of replacing the
Lagrangian $\hat {\cal L}({\cal F})$ by the Routhian ${\cal L}(\lambda)$.

\subsection{\it Linear limit}

Before proceeding with adding the charged membranes to (\ref{genuni}), (\ref{gen4form}), let us briefly discuss the special limit when ${\cal L}$ is a
{\it linear} function of $\lambda$ - i.e. precisely the Henneaux-Teitelboim example, which was augmented with charged tensional membranes in \cite{Kaloper:2022jpv}. 
It is immediately clear that the Legendre transformation (\ref{legtransfgen}) breaks down in this case: if e.g. $\hat {\cal L}({\cal F}) = c {\cal F}$,
$\lambda = c$ and ${\cal L}(\lambda) \equiv 0$. In turn, this is not surprising at all, since as is well known Legendre transformation establishes a relation
between a family of tangents to a curve which is the envelope of this family of straight lines.  When the curve is a straight line, it has a fixed tangent -- i.e. itself -- which is its own envelope.
This shows that the linear example is a degenerate limit of the general case. To see it, we can start with a general $\hat {\cal L}({\cal F})$ and truncate it for simplicity to
only quadratic terms,
\be
\hat {\cal L}({\cal F}) = \frac{\alpha}{2} {\cal F} - \frac{c}{8} {\cal F}^2 + \ldots \, .
\label{trunc} 
\ee
The Legendre transform rules (\ref{legtransfgen}) yield 
\be
\lambda 
= \frac{\alpha}{2} - \frac{c}{4} {\cal F} + \ldots \, , ~~~~~~~~~~ {\cal L}({\lambda})  
= \frac{2}{c} \bigl(\lambda-\frac{\alpha}{2})^2 + \ldots \, .
\label{legtransftrunc}
\ee
In the first equation of (\ref{legtransftrunc}), the $(\ldots)$ terms depend on higher 
powers of $c$, while in the second, the $(\ldots)$ terms depend on higher powers of $1/c$. Clearly,
the limit $c \rightarrow 0$, which reduces (\ref{trunc}) to a linear term only is singular. 

Yet note that without charges a field redefinition exists (as remarked earlier), 
\be
\tilde \lambda = {\cal L}(\lambda) \, , ~~~~~~~~~~~ \tilde {\cal A}_{\nu\lambda\sigma} = \frac{1}{\partial_\lambda{\cal L}(\lambda)} {\cal A}_{\nu\lambda\sigma} \, , 
\label{fredef}
\ee
which completely removes the nonlinearities, since the functional Jacobian of this redefinition is unity,
\be
{\cal J} = \det \frac{\partial (\tilde \lambda, \tilde {\cal A}_{\nu\lambda\sigma}) }{\partial (\lambda, {\cal A}_{\nu\lambda\sigma})}  
=  \det \begin{pmatrix}
\partial_\lambda{\cal L}(\lambda) &0 \\
- \frac{ {\cal A}_{\nu\lambda\sigma} }{\partial_\lambda^2 {\cal L}(\lambda) } & \frac{ 1 }{\partial_\lambda {\cal L}(\lambda) } 
\end{pmatrix}   = 1 \, .
\label{jacobian}
\ee
Although this shows that the singularity $c \rightarrow 0$ can be sidestepped, it also points that the procedure is noncommutative since 
if we took the limit $c \rightarrow 0$ before dualizing, it would be unclear  how to trade variables. This is because the steps involve 
formally divergent contributions to the cosmological constant which should be regulated and subtracted
away. This problem was circumvented in \cite{Kaloper:2022oqv,Kaloper:2022utc,Kaloper:2022jpv},
by promoting $\mpl^2$ into a flux of a second $4$-form and dualizing the theory using bilinear terms. 

Note however that when charged membranes are included, they
obstruct the field redefinition (\ref{fredef}), because the couplings change\footnote{Analogously to the 
change of basis from interaction to propagation eisgenstates in the theory of flavor oscillations.}. 
Hence with membranes included, purely linear theory and
higher power theories are physically distinct. 
In any case, in what follows we will mostly work
with the quadratic truncation of ${\cal L}$, which is practically indistinguishable for our purposes whenever the linear
term dominates. 

\section{Charging Up}

So far we have been working with unimodular gravity/$4$-form theory without charged membranes. However,
a theory of $4$-forms is a gauge theory, describing dynamics of membranes charged under ${\cal A}$. What's more,
the `braneless' theory of $4$-forms, in addition to usual local gauge symmetries ${\cal A} \rightarrow {\cal A} +d \omega$
also has generalized global $3$-form symmetry, associated with the `current conservation' 
$d ^{\ast}{\cal F}=0$ \cite{Gaiotto:2014kfa}. The lore that QG does not permit global 
symmetries \cite{Banks:1991mb,Banks:2010zn} (see also \cite{Reece:2023czb} for a review) is simply incorporated 
by breaking the generalized global symmetry by adding objects charged under ${\cal A}$ -- the 
membranes, precisely. For our purposes, in the semiclassical limit, this means that we should enhance (\ref{genuni}), (\ref{gen4form})
with the inclusion of membrane contributions and boundary terms required to properly provide junction
conditions across the membrane walls. Working with the $\lambda$ variables for convenience, the action becomes 
\ba
S &=& \int d^4x \Bigl\{\sqrt{g} \Bigl(\frac{\mpl^2}{2} R 
- {\cal L}_{\tt QFT} - {\cal L}(\lambda) \Bigr) + 
\frac{1}{3} {\epsilon^{\mu\nu\lambda\sigma}} \partial_\mu \Bigl(\lambda \Bigr) {\cal A}_{\nu\lambda\sigma} \Bigr\} \nonumber \\
&-&  \int d^3 \xi \sqrt{\gamma} \mpl^2[ K ] - {\cal T}_{\cal A} \int d^3 \xi \sqrt{\gamma}_{\cal A} - {\cal Q}_{\cal A} \int {\cal A}  \, .
\label{actionnewmemdcharg} 
\ea
where ${\cal T}_{\cal A}$ and ${\cal Q}_{\cal A}$ are the membrane tension and charge, respectively, 
the term $\propto K$ is the Israel-Gibbons-Hawking term 
for gravity which encodes boundary conditions across membrane walls, and $[...]$ is the jump across a membrane. The coordinates
$\xi$ are coordinates along membrane worldvolume, embedding it in spacetime. The charge terms are 
\be
\int {\cal A} = \frac16 \int d^3 \xi {\cal A}_{\mu\nu\lambda} \frac{\p x^\mu}{\p \xi^\alpha} \frac{\p x^\nu}{\p \xi^\beta} 
\frac{\p x^\lambda}{\p \xi^\gamma} \epsilon^{\alpha\beta\gamma} \, .
\ee
As is customary, ${\cal T}_{\cal A} > 0$ to avoid problems with negative energies. 

Classically, membranes can be added to the spacetime as sources of the gauge fields. Their number is fixed, and the location can change by 
their interactions among themselves, and also with other sources of gravitational fields.  
Quantum mechanically, however, the membranes can nucleate in background fields \cite{Brown:1987dd,Brown:1988kg}. 
Such processes change the distribution of sources and their number, and lead to nontrivial transitions in the background geometry. 
The classical `superselection sectors' (with a fixed number of membranes) now all mix up.  

This induces the evolution in the space of geometries due to 
the variation of $\lambda$. In particular, such transitions change the local value of the cosmological `constant', which unlike in
chargeless unimodular gravity is not a constant anymore. It can change discretely by quantum 
creation of membranes. 
Thus, if we start with a given de Sitter space with an initial value of cosmological constant, it will evolve 
by membrane production into a geometry which is locally de Sitter, but where the cosmological constant varies 
from bubble to bubble. This will look like the original ``old inflation" of Guth \cite{Guth:1980zm}, but with a huge number of 
discretely separated false vacua. 

We have initiated a study of those phenomena in \cite{Kaloper:2022oqv,Kaloper:2022utc,Kaloper:2022jpv} for the special 
case of linear ${\cal L}(\lambda) = \mpl^2 \lambda$. Here, we will redo this analysis for the general case, and extract the salient features
of such `unimodular', or pancosmic, landscapes.

\section{de Sitter Instability}

The quantum membrane discharge in the semiclassical limit can be described by the action (\ref{actionnewmemdcharg}) in Euclidean time. This action 
controls the nucleation processes, and sets their rates, $\Gamma \sim e^{-S_E}$ \cite{Coleman:1977py,Callan:1977pt,Coleman:1980aw}.  Here we follow the
same steps as in \cite{Kaloper:2022oqv,Kaloper:2022utc,Kaloper:2022jpv}: first, we Wick-rotate the action using $t = - i x^0_E$, which gives 
$- i \int d^4x \sqrt{g} {\cal L}_{\tt QFT} = - \int d^4x_E \sqrt{g} {\cal L}^E_{{\tt QFT}}$. Next, with the convention 
${\cal A}_{0 jk} = {\cal A}^{E}_{0jk}$, ${\cal A}_{jkl} =  {\cal A}^{E}_{jkl}$ we get ${\cal F}_{\mu\nu\lambda\sigma} = {\cal F}^{E}_{\mu\nu\lambda\sigma}$. 
In addition $\epsilon_{0ijk} = \epsilon^{E}_{0ijk}$ and $\epsilon^{0ijk} = -\epsilon_E^{0ijk}$. The tension and charge terms transform to
$- i {\cal T}_{\cal A} \int d^3 \xi \sqrt{\gamma} = - {\cal T}_{\cal A} \int d^3 \xi_E \sqrt{\gamma}$ and $i {\cal Q}_{\cal A} \int {\cal A}_i = - {\cal Q}_{\cal A} \int {\cal A}_i$. 
The scalars do not change (but if they include time derivatives, those terms change accordingly). We define Euclidean action by $i S = - S_E$. 
Our Euclidean action is 
\ba
S_E&=&\int d^4x_E \Bigl\{\sqrt{g} \Bigl(-\frac{\mpl^2}{2} R_E + {\cal L}(\lambda) 
+ \Lambda_{\tt QFT} \Bigr) + \frac{1}{3} {\epsilon^{\mu\nu\lambda\sigma}_E} \partial_\mu \Bigl( \lambda \Bigr) {\cal A}^E_{\nu\lambda\sigma} \nonumber \\
\label{actionnewmemeu}
&+&\int d^3 \xi \sqrt{\gamma} \mpl^2[ K_E ]  + {\cal T}_{\cal A} \int d^3 \xi_E \sqrt{\gamma}_{\cal A} - \frac{{\cal Q}_{\cal A}}{6} \int d^3 \xi_E \, {\cal A}^E_{\mu\nu\lambda} \, 
\frac{\p x^\mu}{\p \xi^\alpha} \frac{\p x^\nu}{\p \xi^\beta} 
\frac{\p x^\lambda}{\p \xi^\gamma} \epsilon_E^{\alpha\beta\gamma} \, .
\ea
From here on we drop the index ${E}$. 

We then restrict our attention to `vacuum evolution': we only consider transitions between  
locally maximally symmetric backgrounds. Those have local $O(4)$ symmetry
and so dominate in semiclassical limit since they have minimal Euclidean action \cite{Coleman:1977py,Callan:1977pt,Coleman:1980aw}. 
Therefore we set $\langle {\cal L}^E_{\tt QFT} \rangle = \Lambda_{\tt QFT}$, 
with $\Lambda_{\tt QFT}$ the regulated matter sector vacuum energy to an arbitrary loop order. We can further imagine that the divergent parts 
in the limit where regulator decouples are subtracted away by the counterterm ${\cal L}(\lambda)$, whose finite part is
still completely arbitrary. From the QFT/gravity couplings, we infer, as before  
\cite{Kaloper:2022oqv,Kaloper:2022utc,Kaloper:2022jpv}, that $\Lambda_{\tt QFT} = {\cal M}_{\tt UV}^4 
+ \ldots \equiv \mps  \lambda_{\tt QFT}$, where
${\cal M}_{\tt UV}^4$ is the QFT UV cutoff and ellipsis denote subleading terms \cite{Englert:1975wj,Arkani-Hamed:2000hpr}. 
This means that we can write the total cosmological constant in any patch as
\be
\Lambda_{\tt total} = \Lambda_{\tt QFT} + {\cal L}(\lambda)  = \mpl^2 \lambda_{\tt QFT} + {\cal L}(\lambda) \, ,
\label{totalcc}
\ee
where $\lambda$ can vary from patch to patch across membrane walls.

The transitions induced by nucleations of a single membrane, which are dominant channels here, can be approximated by 
geometries with $\Lambda_{{\tt total} ~ out/in}$ glued together along a membrane, 
where the subscripts {\it out/in} denote parent and offspring geometries (exterior and interior of 
a membrane, respectively). Both of the {\it out/in} geometries are described with the metrics
\be
ds^2_E =  dr^2 + a^2(r) \, d\Omega_3 \, ,
\label{metricsmax}
\ee
where $d\Omega_3$ is the line element on a unit $S^3$. The Euclidean 
scale factor $a$ is the solution of the Euclidean ``Friedmann equation", 
\be
3 \mpl^2 \Bigl( \bigl(\frac{a'}{a}\bigr)^2 - \frac{1}{a^2} \Bigr) = - \Lambda_{\tt total} \, ,
\label{fried}
\ee
which follows because the bulk 
metric-dependent part of (\ref{actionnewmemeu}) is structurally the same as in standard General Relativity. 
The prime designates an  $r$-derivative. From here on, we will drop the subscript ``${\tt total}$". 

To construct these geometries, we need to assemble together two patches, each with a local metrics given by
(\ref{metricsmax}) but with different $\Lambda$, and then use 
the junction conditions to connect the patches. The boundary conditions induced on a membrane 
follow from  (\ref{actionnewmemeu}) by varying with respect to ${\cal A}$. Similarly, the boundary conditions on the
metric follow from Israel-Gibbon-Hawking junction conditions. Summarizing \cite{Kaloper:2022oqv,Kaloper:2022utc,Kaloper:2022jpv}, 
\be
a_{out} = a_{in} \, , ~~~~~~ \lambda_{out} - \lambda_{in}  = \frac{{\cal Q}_{\cal A}}{2} \, ,  ~~~~~~ \mps \Bigl(\frac{a_{out}'}{a} - \frac{a_{in}'}{a} \Bigr)
 = -\frac{{\cal T}_{\cal A}}{2} \, ,
 \label{metricjc}
\ee 
in the coordinate system where the outward membrane normal vector is oriented in the direction of the 
radial coordinate; $r$ measures the distance in this direction. 
The discontinuities in $\lambda$ and $a'$ follow since a membrane is a Dirac $\delta$-function source of charge and tension. 

The next step is to solve (\ref{fried}) for $a' = \zeta_j \sqrt{1-  \frac{\Lambda a^2}{3 \mpl^2}}$, with 
$\zeta_j = \pm 1$ designating the two possible branches of the square root, and rewrite the discontinuous junction conditions 
in a more convenient form, as in \cite{Kaloper:2022oqv,Kaloper:2022utc,Kaloper:2022jpv}.
Using $a'^2_{out} - a'^2_{in} =- a^2  (\Lambda_{out}-\Lambda_{in})/3\mps$ 
which follows from (\ref{fried}), and the last equation of (\ref{metricjc}), we can extract an equation for $a'_{out} + a'_{in}$. Then,
adding and subtracting those two equations, we solve for $a'_{out/in}$. Finally, we solve for $a'_{out/in}$ from (\ref{fried}), 
to obtain 
\ba
\zeta_{out} \sqrt{ 1-  \frac{\Lambda_{out} a^2}{3 \mps}} 
&=& - \frac{{\cal T}_{\cal A}}{4\mps}\Bigl(1 -  \frac{4\mps}{3 {\cal T}^2_{\cal A}} \bigl(\Lambda_{out} - \Lambda_{in} \bigr)\Bigr)\, a \, , \nonumber \\
\zeta_{in} \sqrt{1-  \frac{\Lambda_{in} a^2}{3 \mps}} 
&=& \frac{{\cal T}_{\cal A}}{4\mps}\Bigl(1 +  \frac{4\mps}{3 {\cal T}^2_{\cal A}} \bigl(\Lambda_{out} - \Lambda_{in} \bigr)\Bigr) \, a \, . 
\label{diffroots}
\ea
Here $\zeta_i = \pm$ can be viewed as discrete conserved charges. They pick one of two possible branches of the square root of $\bigl(\frac{a'}{a}\bigr)^2 - \frac{1}{a^2} 
= - \Lambda/3 \mpl^2$. The configurations which result from gluing together sections of exterior and 
interior metrics (\ref{metricsmax}) are therefore counted by the variations of the sign of $\Lambda$ and the 
branches of solutions ($\zeta_j = \pm 1$) of Euclidean Friedmann equation (\ref{fried}). The taxonomy of allowed solutions is  
presented in detail in \cite{Kaloper:2022oqv,Kaloper:2022utc,Kaloper:2022jpv}, and follows the steps described by 
\cite{Brown:1987dd,Brown:1988kg}. Here we focus on the novel ingredients that follow from the generalized theory (\ref{actionnewmemdcharg}). 

The main new ingredient is that, from (\ref{totalcc}), $\Lambda_{out} - \Lambda_{in} = {\cal L}(\lambda_{out}) - {\cal L}(\lambda_{in})$. Combining
this with our last remaining junction condition from (\ref{metricjc}),  $\lambda_{out} - \lambda_{in}  = \frac{{\cal Q}_{\cal A}}{2}$, yields 
\be
\Lambda_{out} - \Lambda_{in} =  {\cal L}(\lambda_{out} ) - {\cal L}(\lambda_{out} - \frac{{\cal Q}_{\cal A}}{2}) \, .
\label{countertermjump}
\ee

This equation has important consequences. To see this, first off, we can check that in the linear limit, ${\cal L} = \mps \lambda$, this yields 
$\Lambda_{out} - \Lambda_{in} = \frac{\mps{\cal Q}_{\cal A}}{2}$. Plugging this into Eqs. (\ref{diffroots}) precisely reproduces the
formulas found in \cite{Kaloper:2022oqv,Kaloper:2022utc,Kaloper:2022jpv}, since the right hand side (RHS) of (\ref{diffroots}) becomes
$\mp\frac{{\cal T}_{\cal A}}{4\mps}\Bigl(1 \mp q \Bigr)$ where $q = \frac{2\mpl^4 {\cal Q}_{\cal A}}{3 {\cal T}^2_{\cal A}}$. 

When ${\cal L}$ is quadratic, ${\cal L} = 2\lambda^2$, we obtain\footnote{For a 
small ${\cal Q}_{\cal A}$, the difference between $\lambda_{out/in}$ is much smaller than either one, and we can ignore the subscript here} 
$\Lambda_{out} - \Lambda_{in} \simeq 4 \lambda \Delta \lambda = 2 \lambda {\cal Q}_{\cal A}$, and so 
the terms controlling the boundary conditions (\ref{diffroots}) on the RHS are
$\simeq \mp\frac{{\cal T}_{\cal A}}{4\mps}\Bigl(1 \mp  \frac{8\mps \lambda {\cal Q}_{\cal A}}{3 {\cal T}^2_{\cal A}} \Bigr)$. This is precisely the behavior found in 
\cite{Brown:1987dd,Brown:1988kg}, and later also encountered in, e.g., \cite{Bousso:2000xa}. The implications for the dynamics are crucial, since these terms control
the selection rules which allow or prohibit a specific type of instanton which mediates transitions of initial de Sitter.  

\begin{figure}[bth]
    \centering
    \includegraphics[width=6.2cm]{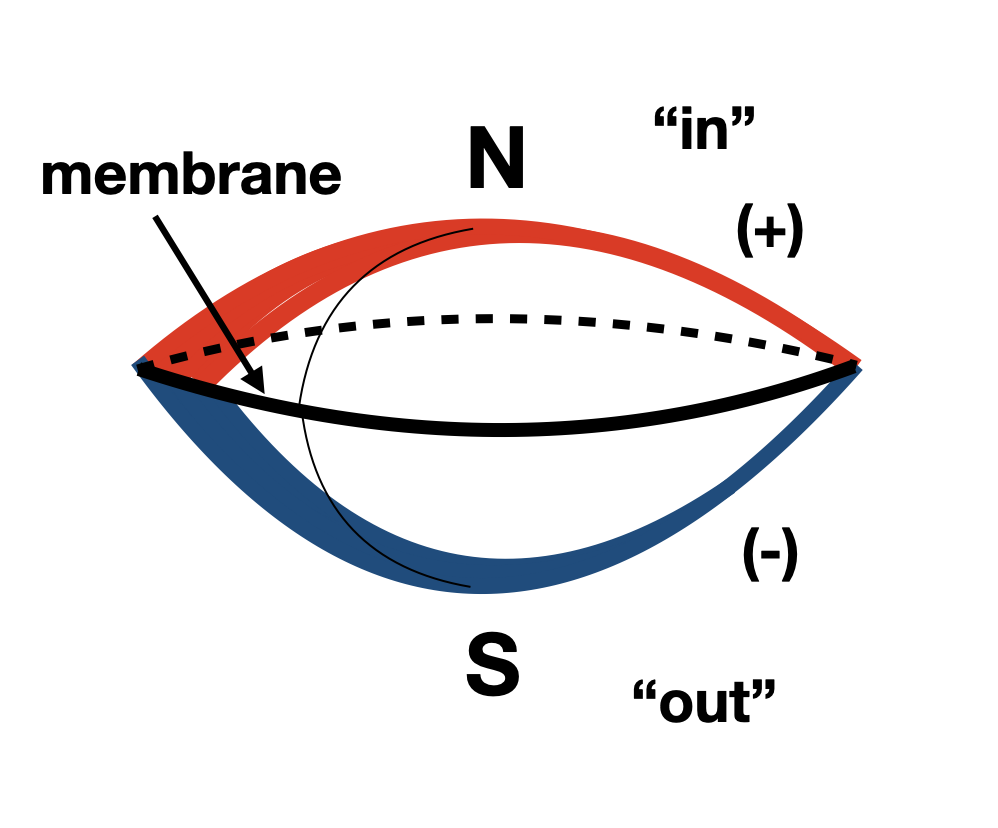}
    \caption{A $|q|<1$ instanton mediating $dS \rightarrow dS$.}
    \label{fig1}
\end{figure}

In the linear case, the point is that for $|q| < 1$ the only allowed instantons are those where the RHS has a fixed sign: $(-)$ in the first and $(+)$ in the second equation.
Since ${\cal T}_{\cal A} > 0$, this uniquely fixes the signs of $\zeta_{out/in}$: $\zeta_{out} = -1, \zeta_{in} = +1$. Thus only instantons which have $(\zeta_{out},\zeta_{in}) = (-,+)$ are allowed.
Specifically, for the transitions of de Sitter to de Sitter, completely independently of the initial and final value of cosmological constant, there is unique allowed process, described 
by the instanton in Fig. (\ref{fig1}); all other possibilities are prohibited \footnote{There is only one more process, whereby $dS \rightarrow AdS$ \cite{Kaloper:2022oqv,Kaloper:2022utc}.}. 

\begin{figure}[thb]
    \centering
    \includegraphics[width=7.2cm]{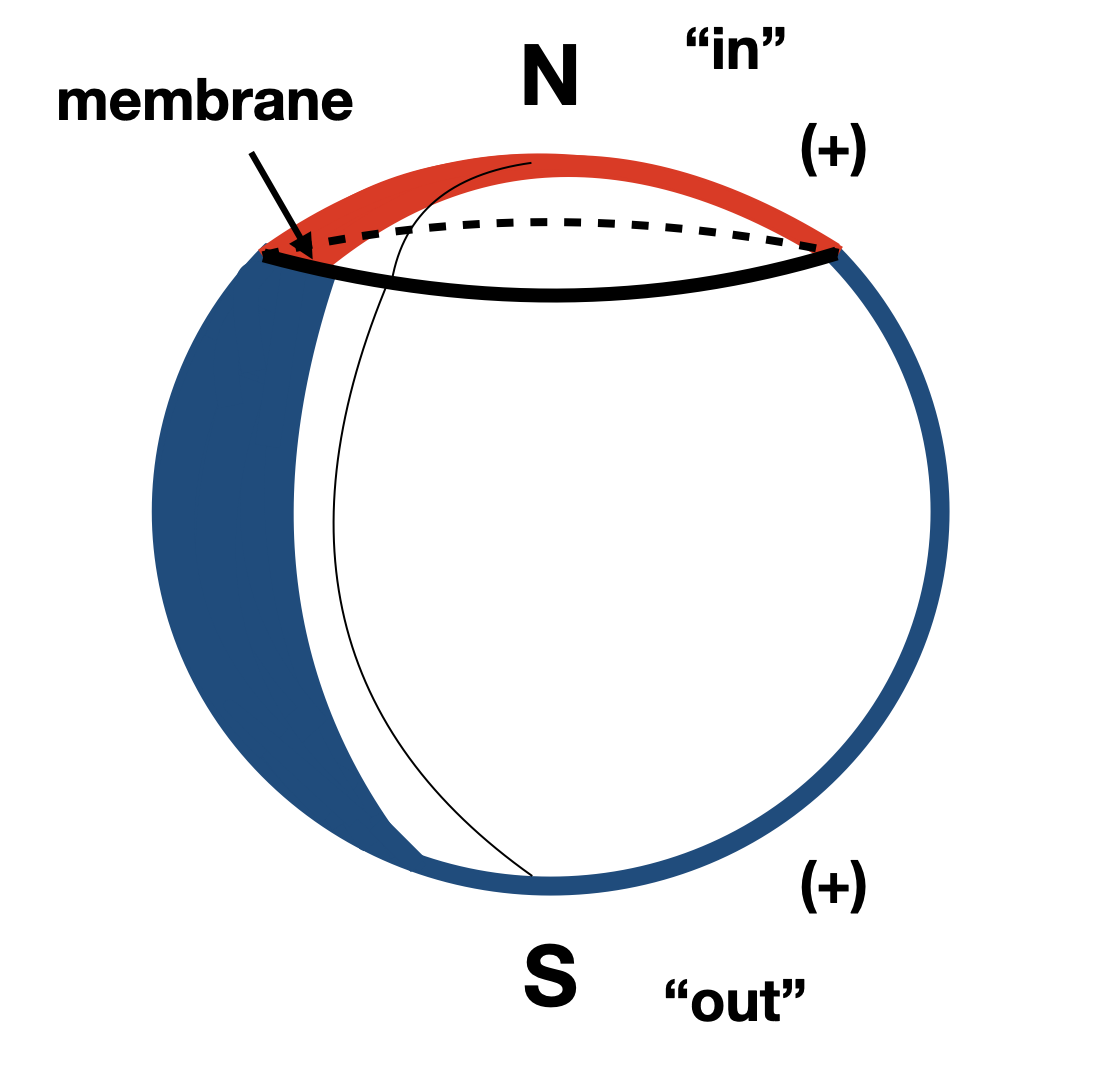}
    \caption{A large flux instanton mediating $dS \rightarrow dS$. }
    \label{fig2}
\end{figure}

In contrast, in the quadratic case, the boundary condition depends on $\Bigl(1 \mp  \frac{8\mps \lambda {\cal Q}_{\cal A}}{3 {\cal T}^2_{\cal A}} \Bigr)$
and therefore on the parent value of the $4$-form flux $\lambda =  - {\bigl(\frac{1}{48 \sqrt{g}} \epsilon^{\mu\nu\lambda\sigma} {\cal F}_{\mu\nu\lambda\sigma}\bigr)}$.
This means that for a fixed ${\cal Q}_A$, when the flux ${\cal F}$ is small the signs on the RHS will again uniquely fix $\zeta_{out}, \zeta_{in}$, allowing only the instanton
of Fig. (\ref{fig1}). However, for large flux, which is generically required to screen a large QFT contribution to cosmological constant $\Lambda_{\tt QFT}$ \cite{Bousso:2000xa},
the situation is reversed: the instanton of Fig. (\ref{fig1}) is disallowed, since the large ${\cal F}$ limit corresponds to $|q|>1$. Instead, only  $(\zeta_{out}, \zeta_{in})=(+,+)$ or $(-,-)$ 
can occur. The dominant instanton in this case is given in Fig. (\ref{fig2}).
The reason this instanton dominates follows from the membrane nucleation rates, 
$\Gamma \sim e^{-S_{bounce}}$ \cite{Coleman:1977py,Callan:1977pt,Coleman:1980aw}. 
To compute them, we need the bounce action, which is defined by 
$S({\tt bounce}) = S({\tt instanton}) - S({\tt parent})$. A precise calculation then yields \cite{Brown:1987dd,Brown:1988kg,Kaloper:2022oqv,Kaloper:2022utc}
\ba
S({\tt bounce}, a) &=& 2 \pi^2 a^3 {\cal T}_{\cal A} - 2\pi^2 \Bigl( \Lambda_{in}\int_{North~Pole}^{a} da \Bigl(\frac{a^3}{a'} ~ \Bigr)_{in}  
+ 3 \mps a^2 
\zeta_{in} \sqrt{1 - \frac{\Lambda_{in} a^2}{3\mps}} ~ \Bigr) 
 \nonumber \\ 
&+& 2\pi^2 \Bigl( \Lambda_{out}\int_{North~Pole}^{a} da \Bigl(\frac{a^3}{a'}\Bigr)_{out}  
+ 3 \mps a^2  \zeta_{out} \sqrt{1 - \frac{\Lambda_{out} a^2}{3\mps}} ~ \Bigr) \, . ~~~
\label{bounce1dact}
\ea
In this formula, the ``outside" contributions to the integrals coming from the parent geometry completely cancel between the instanton and the `parent reference' actions.
The only contributions from the parent come from the complement of the outside geometry of the instantons \cite{Coleman:1980aw,Brown:1987dd,Brown:1988kg}. In the
$(+,+)$ instanton, this means that the parent contributions to the bounce action are small, comparable in size to the `offspring' contribution. As a result, the integral
is dominated by the area surrounding the membrane, since the contributions from the polar regions are small. This is in contrast to the $(-,+)$ instanton of Fig (\ref{fig1}),
where the complement of the parent section of the instanton geometry is most of the initial de Sitter, and so the bounce action will be dominated
by the Euclidean de Sitter parent area -- i.e., precisely the Gibbons-Hawking action, as we will show in detail below.

Whenever the $(+,+)$ instanton dominates, as $\Lambda$ becomes smaller, the action for the
$(+,+)$ instantons converges to (see \cite{Coleman:1977py,Callan:1977pt,Coleman:1980aw,Brown:1987dd,Brown:1988kg,Kaloper:2022oqv,Kaloper:2022utc} and many other papers)
\be
S_{\tt bounce} \simeq \frac{27\pi^2}{2} \frac{{\cal T}_{\cal A}^4}{(\Delta \Lambda)^3}\, .
\label{familiarso}
\ee 
Hence, the decay rate saturates as the cosmological constant decreases and as a result, while any initial de Sitter space will steadily evolve  
towards lower cosmological constant values (easily reaching $\Lambda < 0)$, all values of $\Lambda$ will be
approximately equally likely, and, below some critical value, very long lived. 
It thus seems that as long as these instantons dominate, the only way to argue that 
the terminal cosmological constant induced by transitions is small and nonnegative is to invoke anthropics. This is the reason we think the alternative framework where
the only allowed instanton is $(-,+)$ of (\ref{fig1}) is preferred.

This conclusion remains valid for more general cases, when ${\cal L}$ includes more powers of $\lambda$ as in the 
Taylor series (\ref{lambdalag}), ${\cal L}(\lambda) = c_1 \mpl^2 \lambda + 2c_2 \lambda^2 + 4 c_3 \frac{\lambda^3}{\mpl^2} + \ldots$. 
The equations look more involved, but the physical outcome is similar to the two simple limits above: generically, for small fluxes the instanton (\ref{fig1}) is the only one allowed,
whereas for large fluxes the instanton (\ref{fig2}) dominates. This means, that in the absence of the linear term, all theories with higher powers of $4$-form flux
behave like pure ${\cal F}^2$. However, when the linear term is present, it dominates for all sub-Planckian fluxes ${\cal F}/\mps < 1$ (assuming $c_1 \sim {\cal O}(1)$)! 
The simplest case ${\cal L} = c_1 \mps \lambda + 2 c_2 \mps \lambda^2$, which we elaborated upon above, serves as a clear example.
Since $\Lambda_{out} - \Lambda_{in} = \frac{c_1}{2} \mpl^4 {\cal Q}_{\cal A}  + 2 c_2 \mps \lambda {\cal Q}_{\cal A} + {\cal O}({\cal Q}_{\cal A}^2)$, 
the junction conditions (\ref{diffroots}) to leading order become 
\ba
\zeta_{out} \sqrt{ 1-  \frac{\Lambda_{out} a^2}{3 \mps}} 
&=& - \frac{{\cal T}_{\cal A}}{4\mps}\Bigl(1 -  \frac{2\mpl^4 {\cal Q}_{\cal A}}{3 {\cal T}^2_{\cal A}} \bigl({c_1} + 4 c_2 \frac{\lambda}{\mps} \bigr)\Bigr)\, a \, , \nonumber \\
\zeta_{in} \sqrt{1-  \frac{\Lambda_{in} a^2}{3 \mps}} 
&=& \frac{{\cal T}_{\cal A}}{4\mps}\Bigl(1 +  \frac{2\mpl^4  {\cal Q}_{\cal A}}{3 {\cal T}^2_{\cal A}} \bigl({c_1}  + 4 c_2 \frac{\lambda}{\mps})\Bigr) \, a \, . 
\label{diffrootssqlin}
\ea
Hence whenever $c_2 \sim c_1$ and the flux is sub-Planckian, ${\cal \lambda} < \mpl^2 $, the flux contribution to the junction conditions (\ref{diffrootssqlin}) 
is subleading. Ergo, the large sub-Planckian fluxes might only influence the nonperturbative physics selection rules initially, 
at scales close to $\mpl$, but their influence quickly wanes. In natural theories where it is not prohibited, the linear term does all the driving. 

This persists for general ${\cal L}$. After a straightforward manipulation, one finds to linear order in $\Delta \lambda$  
\ba
\zeta_{out} \sqrt{ 1-  \frac{\Lambda_{out} a^2}{3 \mps}} 
&=& - \frac{{\cal T}_{\cal A}}{4\mps}\Bigl(1 -  \frac{2{\cal Q}_{\cal A}}{3 {\cal T}^2_{\cal A}} \frac{\partial{\cal L}(\lambda)}{\partial (\lambda/\mps)} \Bigr)\, a \, , \nonumber \\
\zeta_{in} \sqrt{1-  \frac{\Lambda_{in} a^2}{3 \mps}} 
&=& \frac{{\cal T}_{\cal A}}{4\mps}\Bigl(1 +  \frac{2  {\cal Q}_{\cal A}}{3 {\cal T}^2_{\cal A}} \frac{\partial{\cal L}(\lambda)}{\partial (\lambda/\mps)}\Bigr) \, a \, . 
\label{diffrootsgen}
\ea
This means that as long as 
\be
\Bigl | Q \Bigr | < 1 \, , ~~~~~~~~~~~~~~~
Q  = \frac{2  {\cal Q}_{\cal A}}{3 {\cal T}^2_{\cal A}} \frac{\partial{\cal L}(\lambda)}{\partial (\lambda/\mps)} \, ,  
\label{gencond}
\ee
the only instantons which can facilitate $dS \rightarrow dS$ transitions (in the limit $\Delta \lambda/\lambda \ll 1$) are those of Fig. (\ref{fig1}). 
This equation generalizes $| q | < 1$ and is always satisfied when the linear term dominates\footnote{One may ask how this compares to the 
`weak gravity conjecture'. Transliterating the inequalities of \cite{Arkani-Hamed:2006emk}, WGC requires ${\cal Q}_{\cal A}/{\cal T}_{\cal A} > 1/\mpl$. 
Combining this and  (\ref{gencond}) yields $\frac{1}{\mps} \frac{\partial{\cal L}(\lambda)}{\partial (\lambda)} < \frac32 \frac{{\cal T}_{\cal A}}{\mpl^3}$. Thus for sub-Planckian
tension membranes this will be satisfied if ${\cal L} = c \mpl^2 \lambda$ with $c<\frac{3}{2} \frac{{\cal T}_{\cal A}}{\mpl^3} < \frac32$. 
This might be realized by, e.g. strong dynamics effects and SUSY breaking below the Planck scale \cite{Aurilia:1980xj}. 
Suffice it to say for now, that the bound does not appear prohibitive.}  ${\cal L}$. 

From here on, for simplicity we will assume that (\ref{gencond}) holds and approximate ${\cal L}$ by the linear term. 
To proceed, we can solve Eqs. (\ref{diffroots}) for $a^2$: 
\be
\frac{1}{a^2} = \frac{\Lambda_{out}}{3\mps} + \Bigl(\frac{{\cal T}_{\cal A}}{4 \mps}\Bigr)^2 
\Bigr(1 -Q \Bigr)^2 
= \frac{\Lambda_{in}}{3\mps} + \Bigl(\frac{{\cal T}_{\cal A}}{4 \mps}\Bigr)^2 
\Bigr(1 +Q \Bigr)^2 \, .
\label{radii}
\ee
As noted in \cite{Kaloper:2022oqv,Kaloper:2022utc,Kaloper:2022jpv}, using this we discern  
two regimes of membrane nulceations. When $a^2$ is comparable to 
de Sitter radii, from Eq. (\ref{radii}), $a^2 \sim (1-\frac{\Lambda_j a^2}{3 \mps})^{1/2} \ll 1$
and so the bounce action is approximately 
\be 
S_{\tt bounce} \simeq - \frac{12\pi^2 \mpl^4 \Delta \Lambda}{\Lambda_{out} \Lambda_{in}} \, .
\label{fastbounce}
\ee
In this regime the transitions are fast because $S_{\tt bounce} <0$. The reverse processes increasing $\Lambda$ are suppressed 
because their bounce action is 
the sign-reversed (\ref{fastbounce}) and so they are 
rarer. Every once in a while de Sitter space is given a push up. But mostly it decays to flatter space, decreasing $\Lambda$. This lasts as long as 
$\Lambda_{out} \gg 3 \mpl^2 \Bigl(\frac{{\cal T}_{\cal A}}{4 \mpl^2}\Bigr)^2$. 

Once $\Lambda$ decreases to $\Lambda \la 3 \mpl^2 \Bigl(\frac{{\cal T}_{\cal A}}{4 \mpl^2}\Bigr)^2$, 
the discharge nucleations proceed via production of small bubbles, which have the 
bounce action \cite{Kaloper:2022utc}
\be
S_{\tt bounce}  \simeq \frac{24\pi^2 \mpl^4}{\Lambda_{out}} 
\Bigl(1- \frac{8}{3} \frac{\mpl^2  \Lambda_{out}}{ {\cal T}_{\cal A}^2} \Bigr)\, ,
\label{familiars2}
\ee 
and $S_{\tt bounce} > 0$ because $\Lambda < 3 \mpl^2  \Bigl(\frac{{\cal T}_{\cal A}}{4 \mpl^2}\Bigr)^2$. 
This action is remarkable: as $\Lambda_{out} \rightarrow 0$ the action diverges. 
Perhaps this is not entirely surprising: as $\Lambda$ decreases the 
geometric entropy ${\cal S}_{GH} \simeq \frac{24\pi^2 \mpl^4}{\Lambda_{out}}$
grows, and the ensuing `chaos' takes over. The decay towards Minkowski then simply looks like the enforcement of
the $2^{\rm nd}$ law of thermodynamics. As a result the bubbling rate $\Gamma \sim e^{-S_{bounce}}$
has an essential singularity at $\Lambda_{out} \rightarrow 0$, where the rate goes to zero \cite{Kaloper:2022oqv,Kaloper:2022utc,Kaloper:2022jpv}. 
Hence when $|Q| < 1$, small cosmological constants become very long lived, 
and the closer the geometry gets to a locally Minkowski  space, the more stable it becomes to discharges. If
it ends up with zero cosmological constant, further discharge stops. 

Hence we see that as a result of augmenting unimodular gravity with charged membranes, 
de Sitter spaces of the theory are immediately rendered
unstable. Membranes will nucleate quantum-mechanically, with the dominant trend being the discharge of the cosmological constant 
in their interior. When the membrane charge and tension in {\it any} generalized unimodular gravity satisfy 
$ \Bigl |\frac{2  {\cal Q}_{\cal A}}{3 {\cal T}^2_{\cal A}} \frac{\partial{\cal L}(\lambda)}{\partial (\lambda/\mps)} \Bigr | < 1$
of Eq. (\ref{gencond}), which can be realized, for example, when ${\cal L}$ is dominated by a term liner in $\lambda$ for $|\lambda| < \mps$, 
the discharge can only occur via a single channel controlled by the instanton of Fig. (\ref{fig1}). In this case, 
when the background cosmological constant decreases below $\Lambda_{critical} \simeq 3 \mpl^2 \Bigl(\frac{{\cal T}_{\cal A}}{4 \mpl^2}\Bigr)^2$, 
the subsequent transitions are slow, since the bounce action is approximately given by the parent geometry horizon area
(\ref{familiars2}) \cite{Gibbons:1976ue,Gibbons:1977mu}.
The deluge of bubbles bounded by membranes eventually comes down to a trickle, clearly favoring Minkowski space as an asymptotic attractor.
In the next section we discuss how close to Minkowski this evolution gets, and how it can provide a means to address the cosmological
constant problem. 

\section{Relaxation of the Cosmological Constant}

The discussion above shows that 
\begin{itemize}
\item 1) de Sitter space is unstable in braney unimodular gravity, 
\item 2) the evolutionary trend is
towards  decreasing $\Lambda$ and 
\item 3) when the flux sector is such that
$ \Bigl |\frac{2  {\cal Q}_{\cal A}}{3 {\cal T}^2_{\cal A}} \frac{\partial{\cal L}(\lambda)}{\partial (\lambda/\mps)} \Bigr | < 1$, which is most simply 
realized when ${\cal L}(\lambda)$ is dominated by the linear term, as in  \cite{Kaloper:2022oqv,Kaloper:2022utc,Kaloper:2022jpv},
the Minkowski space $\Lambda/\mpl^4 \rightarrow 0^+$ is the unique attractor. 
\end{itemize}
These evolutionary trends go in the right direction. Nevertheless, this mechanism of cosmological constant relaxation requires additional ingredients
if it is to be employed to address the cosmological constant problem.

The first issue is the question of how close does the evolution get to Minkowski. On the one hand, the observations imply that the `terminal' achievable value of
$\Lambda$ should be in the range $\Lambda/\mpl^4 \la 10^{-120}$ or so. On the other hand, the evolution of the fluxes by the nucleation of membranes 
 shows that variable $\lambda$ is quantized, changing in units of ${\cal Q}_{\cal A}/2$
 \cite{Teitelboim:1985ya,Teitelboim:1985yc}
\be
\lambda = N \frac{{\cal Q}_{\cal A}}{2} \, .
\label{lkquant}
\ee
Here, as we include general operators ${\cal F}^n, n = 2,3, \ldots$ and not only the linear terms, 
we have set the possible ``background value" $\lambda_0$ to zero imposing 
the Dirac quantization condition \cite{Teitelboim:1985ya,Teitelboim:1985yc} amounting to the addition of 
magnetic monopole sources. Previously, working with strictly 
linear  theory (in the action) in fluxes \cite{Kaloper:2022oqv,Kaloper:2022utc,Kaloper:2022jpv}, we kept $\lambda_0$ explicitly, but since we could absorb it into the cosmological constant counterterm, it was of no consequence. This means, the total cosmological constant in the theory is 
\be
\Lambda_{\tt total} = \mpl^2 \lambda_{\tt QFT} + {\cal L}(N \frac{{\cal Q}_{\cal A}}{2}) = 
\mpl^2 \Bigl(\lambda_{\tt QFT} + N c \frac{{\cal Q}_{\cal A}}{2} + \ldots \Bigr)  \, ,
\ee
where $\lambda_{\tt QFT}$ includes all the contributions to $\Lambda_{\tt total}$ which are not sourced by membranes. The last equality follows
since we are assuming that the linear term dominates in ${\cal L}$. 
Ergo, since the membrane nucleations can only change $\Lambda_{\tt total}$ in the units of ${\cal Q}_{\cal A}/2$, it means that 
a sequence of bubbles can evolve an initial $\Lambda_{\tt total}$ to a terminal value $\Lambda/\mpl^4 \la 10^{-120}$ only if either 
a) the initial value is tuned to be $10^{-120} \mpl^4 + N_{initial} \, c \mps \frac{{\cal Q}_{\cal A}}{2}$ or b) $c \mps \frac{{\cal Q}_{\cal A}}{2} < 10^{-120} \mpl^4$
\cite{Brown:1987dd,Brown:1988kg,Kaloper:2022oqv,Kaloper:2022utc,Kaloper:2022jpv}. The first option is a clear fine tuning, and the 
second is not only theoretically dubitable, but cosmologically problematic since it leads to the empty universe problem, to be discussed
in the next section. 

\begin{figure}[thb]
    \centering
    \hskip1cm    
    \includegraphics[width=6cm]{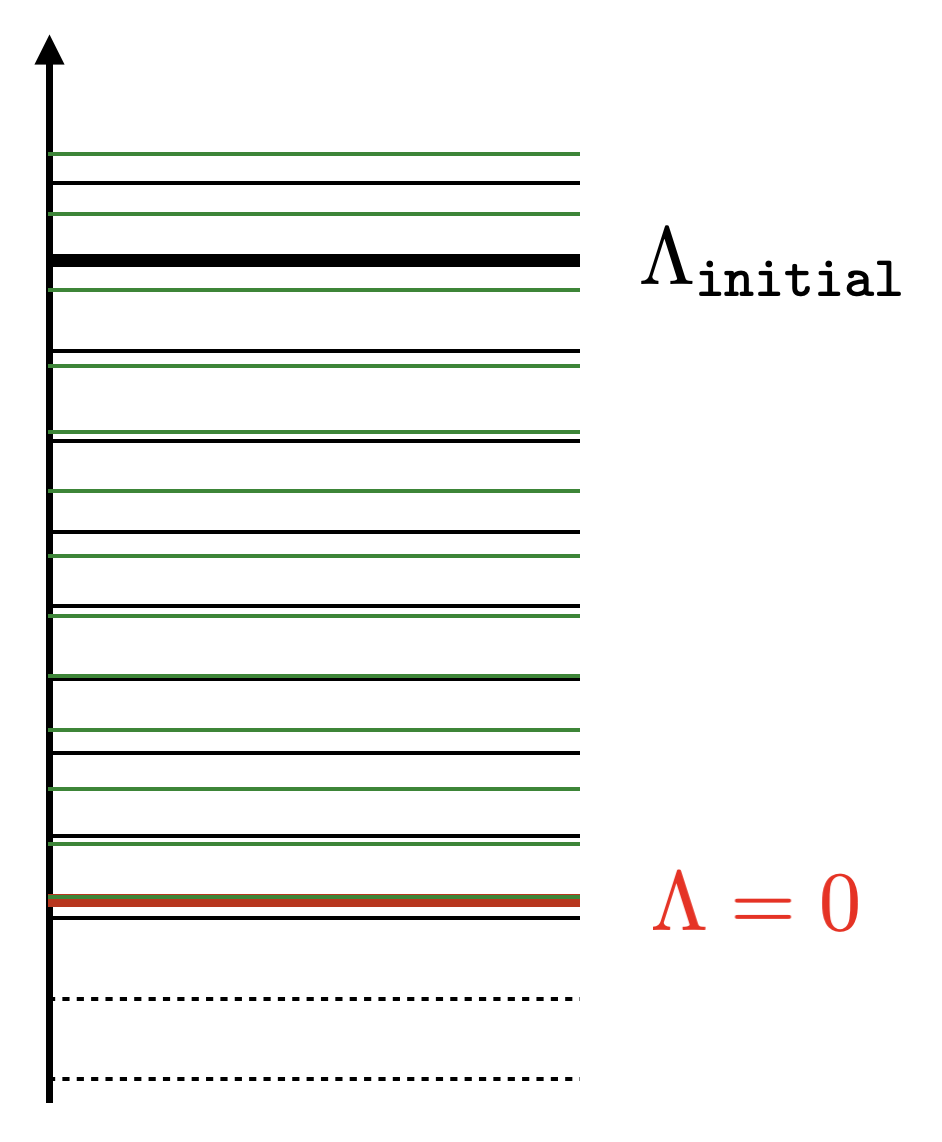}
    \caption{A discharge cascade in the cosmological constant discretuum. Starting with some initial state given by a black line, 
    the evolution will reduce $\Lambda$ to a level where the black and green lines are degenerate, and subsequent evolution 
    can proceed on the green ladder with the terminal step coming arbitrarily close to zero.}
    \label{fig3}
\end{figure}

To get around the fine tuning problem, in \cite{Kaloper:2022oqv,Kaloper:2022utc,Kaloper:2022jpv} a second system of forms and membranes was
introduced, which looked exactly the same as the theory of ${\cal A}$ and its membranes. Since the second form sector is completely degenerate with
the first one,  the total cosmological constant is now\footnote{In  \cite{Kaloper:2022oqv,Kaloper:2022utc,Kaloper:2022jpv} we had $c = \hat c = 1$. Here we keep 
them arbitrary, in principle. }
\be
\Lambda_{\tt total} = \mps \Bigl(\lambda_{\tt QFT} + c \lambda + \hat c \hat \lambda + \ldots \Bigr) = \mpl^2 \Bigl(\lambda_{\tt QFT} + \frac{{\cal Q}_{\cal A}}{2} \bigl( c N + 
\hat c \hat N \omega \bigr) + \ldots \Bigr) \, . 
\label{cctotal}
\ee
Here $\omega = \hat {\cal Q}_{\hat {\cal A}}/{\cal Q}_{\cal A}$ is the ratio of membrane charges. In \cite{Kaloper:2022oqv,Kaloper:2022utc,Kaloper:2022jpv} 
we worked in the linear limit with $c = \hat c = 1$, having renormalized the charges ${\cal Q}_{\cal A}, \hat {\cal Q}_{\hat {\cal A}}$. This means that 
to get arbitrarily close to $\Lambda \rightarrow 0^+$, the charge ratio $\omega$ needs to be an irrational number \cite{Banks:1991mb,niven}. If so, there exist
integers $N, \hat N$ exist such that 
$N + \hat N \omega$ is arbitrarily close to $- \frac{2\lambda_{\tt QFT}^2}{{\cal Q}_{\cal A}}$. The set of values of $\lambda_{\tt QFT} + c \lambda + \hat c \hat \lambda$ 
is discrete but it is dense in a set of reals, with values arbitrarily close to any real number including zero \cite{Banks:1991mb,niven}. Moreover, 
there exist many sequences of discharging membranes, 
which will arrive to $N, \hat N$ at which point the cosmological 
constant is arbitrarily close to zero, and the underlying nearly
flat space is very long lived, due to the pole of the 
bounce action, Eq. (\ref{familiars2}). We illustrate this in Fig (\ref{fig3}). 

To reach this conclusion, what is really required is that the ratio of the energy densities which individual fluxes 
contribute to $\Lambda_{\tt total}$ should be irrational. If charges are mutually irrational, and ${\cal L}(\lambda)$
is dominated by linear terms this follows, but in general other options might occur. The respective flux energies might be
mutually irrational due to compactification effects, if dual $4$-forms arise from dimensional reduction, and/or due to strong coupling
effects in some gauge theory, or thanks to higher powers in ${\cal L}$. E.g. 
$\sqrt{1+x^2} = \sum_k \frac{\Gamma(3/2)}{k! \Gamma(3/2-k)} x^2$ will map integers into irrational numbers.
We will not delve here into the specific details of how irrational variations can arise, deferring that for later work. For the rest of the current paper, we 
will restrict to linear terms, and treat $c {\cal Q}_{{\cal A}}, \hat c  \hat {\cal Q}_{\hat {\cal A}}$ as mutually incommensurate, for simplicity's sake. 

In this case, as Fig. (\ref{fig3}) shows, given that the bounce action (\ref{familiars2}) has a pole at vanishing $\Lambda$, 
the quantum attractor of the evolution, as we will now explain \cite{Kaloper:2022oqv,Kaloper:2022utc,Kaloper:2022jpv}, is 
\be
\frac{\Lambda}{\mpl^4} \rightarrow 0^+ \, .
\ee
Recall the previous approaches with at least quadratic flux contributions to the total cosmological constant 
 \cite{Brown:1987dd,Brown:1988kg,Hawking:1984hk,Duff:1989ah,Duncan:1989ug,Duncan:1990fr,Bousso:2000xa}. 
For a natural cosmological constant $|\Lambda_{\tt QFT}| \sim \mpl^4$,  by Eq. (\ref{totalcc}), this means that ${\cal L} \sim {\cal F}^2 \sim \mpl^4$ and
${\cal F} \sim \mpl^2$. In the pure quadratic theory, this forces $\lambda \sim \mpl^2$. Hence the variable $Q$ of Eq. (\ref{gencond}) which controls which 
instanton the transition goes by is $|Q| \sim \frac {\mpl \lambda}{{\cal T}_{\cal A}} |\frac{\mpl {\cal Q}_{\cal A}}{{\cal T}_{\cal A}}| \gg 1$ for typical
cases where membrane charges and tensions are sub-Planckian\footnote{E.g. in the examples of [36], ${\cal Q}_{\cal A} \sim M_{11}^3/\mpl$ and
${\cal T}_{\cal A} \sim M_{11}^3$ where $M_{11} \ll \mpl$ is the $11D$ Planck scale, and so $|Q| \sim \frac{\mpl \lambda}{M_{11}^3} \gg 1$, as claimed. Note also that  this yields $\frac {{\cal Q}_{\cal A}}{{\cal T} _{\cal A}} \sim \frac{1}{\mpl}$ and so the setup of [36] can marginally satisfy the weak gravity conjecture, as does
the present work (see footnote 12). The recent paper  [64], while generally confirming our observations from 
[5,6] and the present work, claims the opposite.}. Thus
in this case the only transitions which occur are those mediated by the instanton of Fig. (\ref{fig2}), whose bounce action saturates at (\ref{familiarso}). 
The transition rate from one value of $\Lambda$ to another is independent of $\Lambda$, and so all values will be
approximately equally likely. As a result, with a uniform distribution of initial values, the resulting dynamical 
distribution of values of $\Lambda$ will remain uniform, 
since ``anything goes" at the same rate when the instantons of Fig. (\ref{fig2}) are allowed. 
To favor a small positive cosmological constant, 
one then needs to invoke the anthropic principle \cite{Polchinski:2006gy}.
  
The addition of the linear term in ${\cal F}$ changes this picture dramatically \cite{Kaloper:2022jpv,Kaloper:2022utc,Kaloper:2022oqv}. 
When vacuum energy from QFT is $\Lambda_{\tt QFT} < \mpl^4$, the linear term $\mpl^2 \lambda$  
dominates, ensuring $|Q|<1$ and so the only processes which allow subsequent 
discharges of $\Lambda$ are those mediated by the instanton of Fig. (\ref{fig1}). 
Hence when the cosmological constant reduces to below $\Lambda_{critical} \simeq 3 \mpl^2 \Bigl(\frac{{\cal T}_{\cal A}}{4 \mpl^2}\Bigr)^2$, their bounce action is approximated by the exponent of the Gibbons-Hawking action (\ref{familiars2}), and the further decay
progressively slows down; the evolution relaxes $\Lambda$ to $0^+$ by quantum Brownian drift, 
and it exponentially slows down as $\Lambda \rightarrow 0^+$. This means, as long as $\Lambda > 0$ and there are smaller values, eventually regions of 
smaller $\Lambda$ will be created inside the parent region. 
The most stable universes are those with the smallest value of the positive cosmological constant. 

Further, as long as the cosmological constant is not exactly zero,
one further decay to $\Lambda < 0$ or an up-jump to a large positive $\Lambda$ are possible. In the former case, the evolution completely stops. If there is reheating and normal matter in the universe, this region will collapse into a black hole surrounded by the parent geometry. It is a terminal sink. On the other hand,
if there is an up-jump, the sequence of discharges will repeat, seeking to find an ever smaller 
$\Lambda$ once again. This gives the universe with a larger $\Lambda$ yet another chance to discharge it, by going first up and then discharging toward 
$\Lambda \rightarrow 0^+$. The process may be extremely slow, but that doesn't really matter as long as it can occur. 
As a result, the smaller values of $\Lambda$ will also be statistically more likely\footnote{If the bounce action is sufficiently big, a universe with a positive cosmological constant larger than observed may still be very long lived. But that is not the issue: with %a uniform prior distribution of initial values, and 
evolution 
driven by the instantons of Fig. (\ref{fig1}), it will
not be typical. There will be many more longer lived universes which have a smaller cosmological 
constant, and where subsequent cosmology could produce structures regardless of anthropics.}. The posterior distribution of $\Lambda$ including the tunneling
dynamics effects will not be uniform even if the prior (number frequency) distribution is flat. This 
dynamics favors smallest values. 

Of course, to make sure that arbitrarily small values of cosmological constant can be approached, as we noted above there must be at least two sets of membranes, with incommensurate fluxes. In this case the configurations with tiny positive cosmological constant will accumulate, overwhelming other outcomes. Thus finding a tiny positive 
$\Lambda$ is natural without anthropics. Rather interestingly, this can be achieved already with the simple linear + quadratic actions mentioned in \cite{Aurilia:1980xj}. 
Still we stress that this assumes no phase transitions at late times, in the matter sector after inflation, and it does not immediately lead to a prediction of current cosmic acceleration. 

So we see that the likeliest values of the cosmological constant are small as opposed to large, addressing the
``why is cosmological constant not huge?" part of the problem. 
However we would be too hasty to declare victory at this point. Besides the obvious point that observations suggest
that $\frac{\Lambda}{\mpl^4} \sim 10^{-120}$ instead of zero, in our analysis of the evolution we have treated  $\Lambda_{\tt QFT}$ as
a fixed number, ignoring any possible phase transitions in a late universe. This is an issue, since while the early evolution favors
the tiniest values of $\Lambda_{\rm total}$, the membrane nucleation processes which lead to it slow down tremendously when the 
local value of $\Lambda$ becomes smaller than $\Lambda_{critical} \simeq 3 \mpl^2 \Bigl(\frac{{\cal T}_{\cal A}}{4 \mpl^2}\Bigr)^2$. 
Indeed, from a practical point of view, we need to worry about a framework where the early evolution sets $\Lambda_{\tt total}$ to almost zero, and then at a much later
stage a phase transition in the matter sector occurs, changing the QFT vacuum energy dramatically, and by a large value. 
A region of the universe where this
occurs would typically develop a large negative energy density, which would force it to stop expanding and recollapse, presumably at least 
forming a large black hole. Note, that we can solve this problem immediately by going back to anthropics to fix the very late value of the cosmological 
constant \cite{antropat}.

\begin{figure}[thb]
    \centering
    \includegraphics[width=10cm]{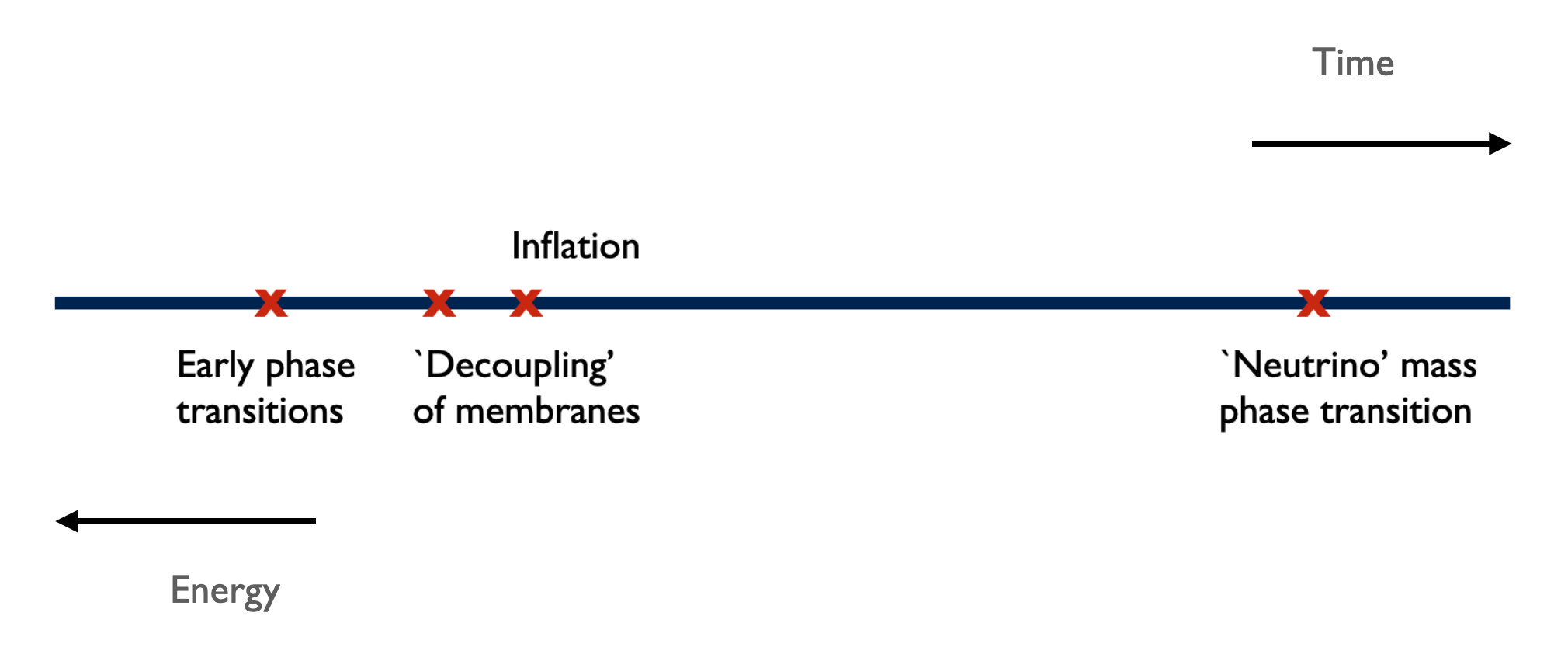}
    \caption{A toy model universe with the cosmological constant problem completely solved. }
    \label{fig4}
\end{figure}

On the other hand, this also points towards a toy model universe where the cosmological constant could be completely solved using our mechanism,
although it is not our universe. Imagine a universe where all but one matter sector phase transitions occur very early, at very high energy 
scales which are above the critical energy density controlling membrane decoupling and the scale of inflation. Quantum dynamics would
favor vacua with nearly vanishing energy density, and after membranes stop being nucleated these 
regions could inflate in slow roll regime, reheat and undergo radiation and matter evolution. If then a new phase transition occurs at a very low
scale, it would typically induce a net negative cosmological constant. However if the theory also contains an ultralight field (an `axion') which 
develops a potential, its potential energy due to misalignment could compensate the negative vacuum energy and make the expansion rate of this region 
start to accelerate while the axion is away from its minimum (see Fig. (\ref{fig4})) \cite{Barbieri:2005gj}. This looks like an approximation to our 
universe, except for the fact that it does not
contain electroweak and QCD phase transition. 

 The discussion above however points to an important difference in how the real cosmological constant problem manifests in our framework, and also other 
 frameworks which strive to address it using dynamical decay -- i.e., adjustment. Some of these points have been already raised in \cite{Polchinski:2006gy}.
 Here we will revisit them, and elaborate some aspects further. First, note that the oft-encountered phrasing of the cosmological constant problem in terms of the
 UV sensitivity of $\Lambda_{\tt total}$ is in fact misleading: the fact that the cosmological constant might get a large contribution from  a very heavy field 
 is only an issue if such a field exists. On the other hand, there are many fields in our universe whose mass scales are much greater than the observed
 scale of vacuum energy. What this shows is that the cosmological constant problem really isn't a UV problem (``why is it not huge?") or an IR problem (``why is it tiny
 but not zero?") per se - it is really an all scales problem, or at least, all scales where we find massive fields (which means at least most of the stuff coming
 from the Standard Model, and possibly from dark matter sectors). In \cite{Polchinski:2006gy}, Polchinski argued compellingly that to get all those contributions
 to conspire to cancel each other down to $10^{-120} \mpl^4$ in local QFT seems well nigh impossible. 
 
 Despite some brave forays in the literature exploring alternatives to local QFT, we share the view of \cite{Polchinski:2006gy}, 
 which adheres to more conventional physics, whereby the cosmological
 constant decays via a de Sitter instability is a more palatable avenue. 
In this approach, regardless of the diversity of contributions from various scales, since they all add to the cosmological constant in the same way thanks to 
Equivalence Principle, they all get screened by bubble nucleation that de Sitter decays by.

\section{Cosmic Connections}

\begin{figure}[bth]
    \centering
    \includegraphics[width=7cm]{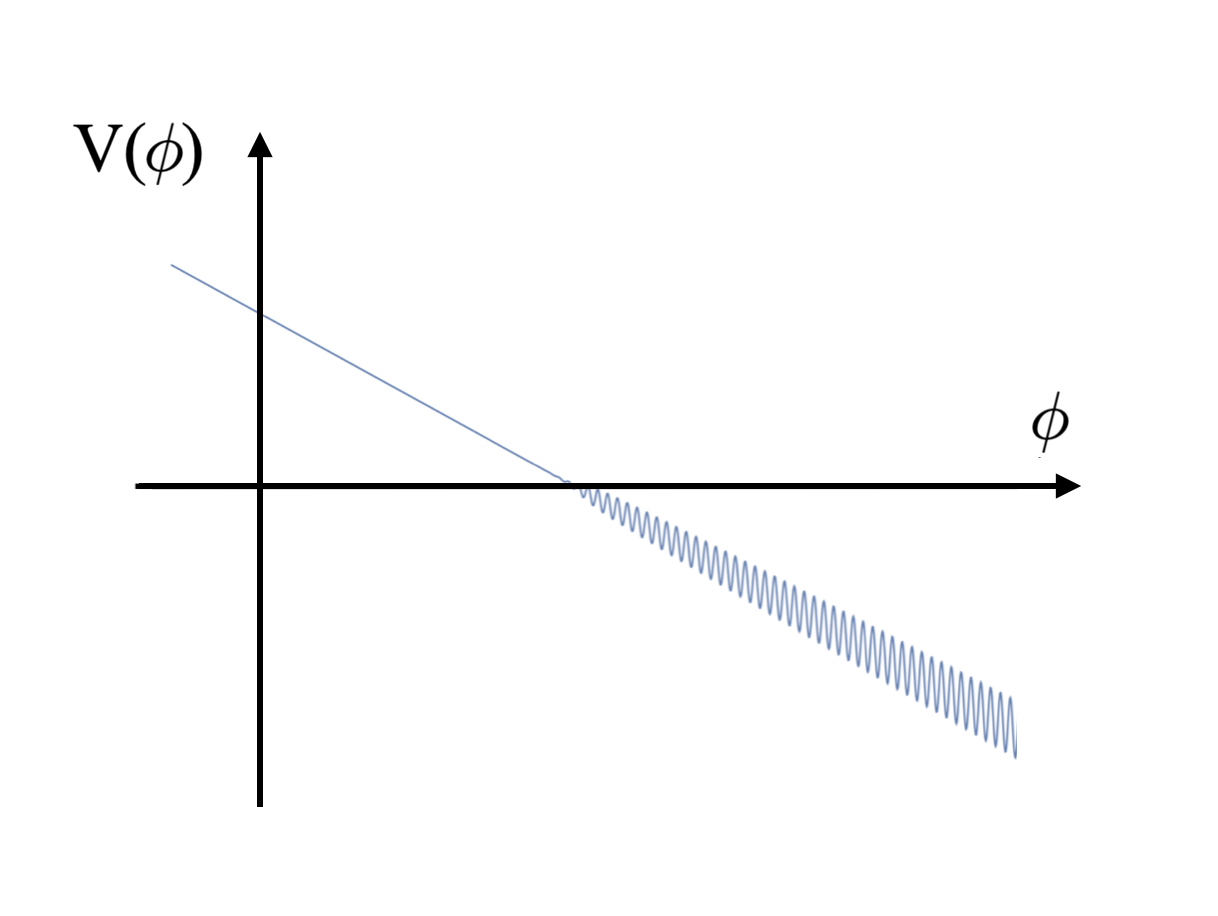}
    \caption{A schematic of Abbott's adjustment mechanism.}
    \label{figa}
\end{figure}

Our mechanism {\it is} a cosmological constant adjustment mechanism, albeit not one mediated by a smooth field, but by a discretely varying flux of a
set of $4$-forms. It shares some features with the adjustment mechanism proposed by Abbott in a very insightful paper \cite{Abbott:1984qf}. 
Abbott proposed a field-theoretic
adjustment mechanism using a scalar with a linear potential which is degenerate with the cosmological constant 
so that the scalar was screening away the cosmological term. Near the zero value of the cosmological constant, Abbott proposed that
strong-coupling nonperturbative corrections arise, like for an axion. They catch the scalar and stop further rolling (see Fig. (\ref{figa})).
However since the adjustment is continuous, the scalar 
must dominate the universe at least until the cosmological constant were nearly zero. In such a universe, inflation does not end
until the Hubble parameter decreases to 
$\simeq 10^{-34} {\rm eV}$, and so the universe never reheats\footnote{This also happens in the adjustment mechanism employing a single $4$-form when
the value of charge ${\cal Q}_{\cal A}$ is selected to be tiny, as noted in \cite{Brown:1987dd,Brown:1988kg}.}.  
In contrast, in our case this empty universe problem is averted since the relaxation of $\Lambda$ 
involves jumps by large charges ${\cal Q}_j$: the cosmological constant jumps by a 
large step $\propto \mps {\cal Q}_j$. The tiny terminal $\Lambda$ arises from the 
misalignment of the large fluxes sourced by charges. As a result, the cosmological constant does not always dominate, but just 
early on \cite{Kaloper:2022oqv,Kaloper:2022utc,Kaloper:2022jpv,Bousso:2000xa}. 
Before the universe jumps to tiny $\Lambda$, it had a large cosmological constant, and
so immediately after its decay it can inflate and reheat using a standard slow roll stage \cite{progress}. 
The discrete adjustment also bypasses Weinberg's no-go theorem \cite{Weinberg:1987dv}. 

In striking contrast to what we'd expect if an adjustment were based on a smooth variation of a field, 
the evolution does not completely wipe out the future of the information about the ancestry of the final near-Minkowski space. If the
adjustment were via a smooth field, the universe everywhere would transition to the final near Minkowski at roughly the same time 
(as Abbott explained that would have to be about now). Instead, discrete adjustment happens locally. In each terminal
bubble the interior geometry will preferentially be nearly flat. However those regions are separated from each other by highly curved spacetime, with large
cosmological constant values, which are still bubbling.

This image of global structure of spacetime is just like the multiverse of eternal inflation (for reviews, see e.g. \cite{Guth:2000ka,Linde:2015edk}). 
After a moment's thought, this is not surprising in the least. In a setup where the net cosmological constant includes a contribution from a 
background flux, and where the $4$-form gauge fields are sourced by charged tensional membranes, the discharge processes analogous
to Schwinger particle production are inevitable \cite{Brown:1987dd,Brown:1988kg}. Since they change the cosmological constant locally, inside
a membrane, given some `initial' parent geometry (by this we mean any patch without bubbles at an arbitrarily selected moment of
observation) it will start to convert into the bubbles of other vacua. However, since the parent is de Sitter, the descendant bubbles will not 
percolate, just like in the original ``old inflation" of Guth \cite{Guth:1980zm}. In most of the bubbles, the cosmological constant will be smaller,
but in some it will be (much!) larger than the parent's \cite{Steinhardt:1982kg,Linde:1986fc}. In those regions inflation will restart itself. In our case, 
since we look at models with two systems of charged membranes with mutually irrational fluxes, there will be a huge number of different vacua, 
since the values of the cosmological constant span a discretuum like in \cite{Bousso:2000xa}, instead of only a few types usually encountered in 
simple models of false vacuum inflation. 

Amusingly, our adjustment mechanism could address a critique directed at inflation in \cite{Turok:1998he}. The authors of this work open up 
with the assertion that ``inflationary theory has for some time had two skeletons in its cupboard". They elaborate that one problem is the selection 
of the initial conditions for the inflaton sector, required to start inflation, and the other is the cosmological constant, which they argue is troublesome 
since inflation happens because it is supported by cosmological constant which in turn must be cancelled. The authors infer that both of `these 
issues point to severe fine tunings. In our picture, both of these problems could be addressed with the help of eternal inflation. 

Indeed, as long as there is even a single region of the universe with $\Lambda > 0$, it can happen that an empty 
universe with a nearly vanishing $\Lambda$ `restarts' itself by a rare quantum jump which increases the cosmological constant 
\cite{Garriga:1997ef,Garriga:2000cv}. In subsequent evolution back to $\Lambda \rightarrow 0$, the repeated process of membrane discharges 
reducing $\Lambda$ back toward zero can scan for an inflationary stage \cite{Garriga:2000cv}. This suggests that the natural `ground state' of the theory
is near-Minkowski, which is a dynamical attractor of the evolution, similar to the proposal of  \cite{Carroll:2004pn,Carroll:2005it} (motivated by the ideas
about reformulating inflationary initial conditions selection). If we follow this path, inflation seems {\it \`a priori} rare. However repeated `recycles' 
allow for filling up the phase space, and so even a `rare' event will be found 
eventually. As Guth points in \cite{Guth:2004tw}, ``once inflation starts, it generically continues forever, creating an infinite number of ``pocket'' universes."
The only difference here is that this eternal selfreproduction does not require a scalar field and its large quantum fluctuations. The large discrete 
variations of the cosmological constant come from the quantized flux which sources vacuum energy.

Guth also stresses the need for the beginning of inflation, due to the past incompleteness theorem \cite{Borde:2001nh}. 
However, as \cite{Guth:2000ka,Carroll:2005it} note, it is possible that some quantum creation event may have initiated the whole process.
We will set this issue aside for now, and only note, that 
it should be possible to embed a standard late cosmology in the present framework. 

Finally, one can consider specific predictions and implications for observations \cite{Freivogel:2005vv}, 
among which might be a past record of colliding with 
other bubbleworlds \cite{kleban,aguirre,Feeney:2010jj}, applications to particle physics hierarchies \cite{Giudice:2019iwl,Kaloper:2019xfj}, 
and maybe even late time variations
of cosmological parameters, as in e.g. \cite{fractal,Marra:2021fvf}. These questions warrant further scrutiny.

\section{Summary}

An established view of General Relativity is that it is a theory of gravity with fixed values of $G_N$ and $\Lambda$. This description is incredibly successful 
for explaining a huge host of observed phenomena, and passes a number of tests and bounds with flying colors. However from the theoretical point of view,
this description runs into a quandary when we ask questions about the reasons why the values of $G_N$ and $\Lambda$ are what they are.

Here, and also earlier, in \cite{Kaloper:2022oqv,Kaloper:2022utc,Kaloper:2022jpv}, we have considered an approach which `liberates' these gravitational 
parameters and allows them to be treated as global gauge invariant degrees of freedom. Focusing on the cosmological constant, this leads to a 
generalized unimodular gravity theory, which is locally completely indistinguishable from GR, 
classically and as a quantum field theory \cite{Fiol:2008vk,deBrito:2021pmw,Kugo:2021bej}.
We showed that this theory is a theory of $4$-forms in disguise, and to make sure that the gauge symmetry is properly encoded, we added a system
of charged tensional membranes for each form present. It then follows that cosmological constant 
is unstable to quantum-mechanical, nonperturbative, discharge 
of membranes. This is because the $4$-form fluxes are degenerate with the cosmological constant by covariance and construction. 
Cosmological constant decays toward $\Lambda/\mpl^4 \rightarrow 0$, which is a dynamical attractor when the theory is dominated by terms linear 
in $4$-form flux, at least in the statistical sense, 
once charge to tension ratio is fixed to (\ref{gencond}). This is an avatar of 
Coleman and De Luccia's `gravitational stabilization' of
flat space to nonperturbative instabilities. This addresses a part of the cosmological constant problem in our universe: it shows why the cosmological 
constant is not huge, without using anthropics.

The post-inflationary phase transitions in the matter sector -- the electroweak and the QCD ones -- obscure the full solution at the moment. It would be interesting to 
explore this issue further. It is also of interest to determine precisely how to connect our mechanism with slow roll inflation, and how to accommodate 
current observed cosmic acceleration (by a very late phase transition?). We plan to return to these questions elsewhere.

\vskip.3cm

{\bf Acknowledgments}: 
We are especially grateful to A. Westphal for collaboration and for many valuable comments, discussions and insights. 
We thank G. D'Amico, S. Dimopoulos, A. Lawrence and J. Terning for valuable comments and discussions. 
NK is supported in part by the DOE Grant DE-SC0009999.

\end{document}